\documentclass[reprint,pra,twocolumn,amsmath,amssymb,superscriptaddress,nofootinbib,showpacs,aps]{revtex4-1}
\pdfoutput=1
\usepackage{graphicx}
\usepackage{bm}
\usepackage{times}
\usepackage{comment}
\usepackage{dsfont}
\usepackage{braket}

\usepackage{color}
\definecolor{blue}{RGB}{0,0,255}
\usepackage[colorlinks,citecolor=blue,linkcolor=blue,urlcolor=blue]{hyperref}

\graphicspath{{"fig/"}}

\begin{document}
\def\BE{\begin{equation}}
\def\EE{\end{equation}}
\def\BY{\begin{eqnarray}}
\def\EY{\end{eqnarray}}
\def\BI{\begin{itemize}}
\def\EI{\end{itemize}}
\def\L{\label}
\def\nn{\nonumber}
\def\({\left (}
\def\){\right)}
\def\[{\left [}
\def\]{\right]}
\def\<{\langle}
\def\>{\rangle}
\def\BA{\begin{array}}
\def\EA{\end{array}}
\def\dsp{\displaystyle}
\def\ds{\displaystyle}
\def\k{\kappa}
\def\dd{\delta}
\def\D{\Delta}
\def\w{\omega}
\def\W{\Omega}
\def\v{\nu}
\def\a{\alpha}
\def\b{\beta}
\def\e{\varepsilon}
\def\d{\partial}
\def\g{\gamma}
\def\G{\Gamma}
\def\tt{\theta}
\def\t{\tau}
\def\s{\sigma}
\def\+{\dag}
\def\8{\infty}
\def\x{\xi}
\def\m{\mu}
\def\l{\lambda}
\def\ii{\textbf{i}}
\def\={\approx}
\def\xc{\frac{2x}{c}}
\def\->{\rightarrow}
\def\r{\vec{r}}
\def\k{\vec{k}}
\def\sinc{\mathrm{sinc}}
\def\xx{\textbf{x}}
\def\yy{\textbf{y}}
\def\qq{\textbf{q}}
\def\rr{\boldsymbol{\rho}}
\newcommand{\ud}{\,\mathrm{d}} 
\def\out{|\textrm{out}\rangle}
\def\inn{|\textrm{in}\rangle}
\def\sqz{|\textrm{sqz}\rangle}
\def\A{\text{A}}
\def\B{\text{B}}
\def\AB{\text{AB}}

\def\length{0.32}

\title{How subtraction of a single photon affects many quantum modes}

\author{Valentin A. Averchenko}
\email{Valentin.Averchenko@mpl.mpg.de}
\affiliation{Max Planck Institute for the Science of Light, G\"{u}nther-Scharowsky-Stra\ss e 1/Building 24, 90158 Erlangen, Germany}              
\affiliation{Institute for Optics, Information and Photonics, University Erlangen-N\"{u}rnberg, Staudtstr.7/B2, 90158 Erlangen, Germany}

\author{Cl\'ement Jacquard}
\affiliation{Laboratoire Kastler Brossel, Universit\'{e} Pierre et Marie Curie-Paris 6, ENS, CNRS; 4 place Jussieu, 75252 Paris, France}

\author{Val\'erian Thiel}
\affiliation{Laboratoire Kastler Brossel, Universit\'{e} Pierre et Marie Curie-Paris 6, ENS, CNRS; 4 place Jussieu, 75252 Paris, France}

\author{Claude Fabre}
\affiliation{Laboratoire Kastler Brossel, Universit\'{e} Pierre et Marie Curie-Paris 6, ENS, CNRS; 4 place Jussieu, 75252 Paris, France}

\author{Nicolas Treps}
\affiliation{Laboratoire Kastler Brossel, Universit\'{e} Pierre et Marie Curie-Paris 6, ENS, CNRS; 4 place Jussieu, 75252 Paris, France}

\date{\today}

\begin{abstract}
The subtraction of a single photon from a multimode quantum field is analyzed as the conditional evolution of an open quantum system. 
We develop a theory describing different subtraction schemes in a unified approach and we introduce the concept of subtraction modes intrinsic to the process.
The matching between the subtraction modes and the modes of the field defines different possible scenarios for the photon subtraction.
In particular, our framework identifies: the conditions of pure photon subtraction, the quantum states of the field modes conditioned to the photon subtraction, the mode with the highest fidelity with a single-photon state when the subtraction is performed on multimode squeezed light.
We use our theory to analyze the photon subtraction from a highly multimode quantum resource - a train of quantum squeezed or correlated optical pulses.
Performing the photon subtraction optimally on various multimode light field has the potential to implement a number of quantum information protocols in a multiplexed and scalable way.
\end{abstract}


\maketitle

\section{Introduction}

In the context of universal quantum computation, all-optical setups have proved to be solid candidates mostly for their robustness against decoherence.
It is commonly accepted that two necessary elements are required: quantum states with negative Wigner function and multipartite entanglement.
In the discrete variable regime of quantum optics, single-photons are generated probabilistically and possess an intrinsic negative Wigner function.
One then has to combine them to produce multipartite entanglement.
Such an approach is actively being explored and led to promising results \cite{OBrien2003}.
On the other hand, in the continuous variable regime, high multipartite entanglement can be easily generated deterministically with Gaussian quantum states \cite{Roslund2013}.
One then has to perform a single conditional non-Gaussian operation to turn the initial resource into a worthy candidate for quantum computation.
The most straightforward non-Gaussian process one can think about are single-photon subtraction and addition.

Single-photon subtraction has been commonly used to perform many optical tasks in the continuous-variable regime.
The successful realizations using this technique include the generation of various key quantum states for quantum optics such as Fock states \cite{Nielsen2007, Yukawa2013a} and cat-like states \cite{Ourjoumtsev2006}.
In a similar fashion, single-photon subtraction is used in quantum state engineering to achieve hybrid entanglement \cite{Morin2014} and to purify entanglement between parties through entanglement distillation \cite{Ourjoumtsev2009,Takahashi2010}.
More fundamentally, it has allowed to probe quantum commutation rules along with single-photon addition \cite{Parigi2007}.
In the perspective of quantum computation, single-photon subtraction is meant to allow one to turn a Gaussian state into a non-Gaussian state thus implementing universal non-Gaussian gates such as a cubic gate \cite{Marek2011,Yukawa2013b,Marshall2014}.
Also, it was demonstrated more recently, that an assembly of photon subtracted quantum states are suitable to tackle the boson sampling problem and its intrinsic complexity \cite{Olson2015}.

Nevertheless, the subtraction of a single photon has rarely been mixed with some high multipartite entanglement in an experiment.
Its theoretical description has been extensively studied in a modal approach on a single mode resource \cite{Tualle-Brouri2009} and a two modes multimode resource \cite{Takahashi2008,Ourjoumtsev2009,Morin2014}, the general multimode theory has never been developed.
Nowadays, highly multimode light fields are produced by a variety of setups, especially in the continuous variable regime where many modes hosting squeezed vacuum states are mixed together \cite{Su2007,Yukawa2008,Roslund2013,MedeirosdeAraujo2014,Chen2014}.
With single mode resource, the challenge consists in matching the single-photon detection mode to the mode of interest as a mismatch mixes the original state with vacuum and results in the decoherence of the quantum resource.
When the resource is multimode, the subtraction or the addition of a single photon happens in a mode-selective manner thus paving the way to the creation of multimode non-Gaussian states.

In the following, we model the single-photon subtraction of a multimode quantum state as the combination of light splitting with a multimode beam splitter and single-photon detection.
We show in Sec. \ref{general} that the two processes combine into an abstract apparatus that performs single-photon subtraction onto a set of eigenmodes with associated subtraction probabilities.
With this general framework, we compute the density matrix of a multimode quantum state after a conditional single-photon subtraction and we show how the subtraction modes naturally appear.
We discuss the issue of matching the subtraction eigenmodes to the modes of the input state in Sec. \ref{sec:single_multi} and assess the dependence of the output state purity upon the modes of the subtraction procedure.
We show that the single-photon subtraction can be pure or non-pure and characterized by a Schmidt number \cite{Ekert1995}.
We then consider multimode squeezed vacuum as an input state in Sec. \ref{sec:squeezed_light} to illustrate the aforementioned concept of matching and derive two figures of merit: the purity of the subtracted multimode state and the purity of the state embedded within a single mode.
Also, in the limit of weak squeezing, single-photon subtraction heralds a state similar to a single-photon state and we construct the optical mode that maximizes the fidelity.
In Sec. \ref{sec:time-domain}, we apply our formalism to the particular case of the single-photon subtraction from a temporally/spectrally multimode light.
We derive the subtraction eigenmodes and their probabilities by precisely describing the single-photon detection and two light splitting mechanisms.
The first one is linear and relies on a weak beamsplitter and a spectral filter \cite{Dakna1996}, the second one is non-linear and based on sum-frequency generation \cite{Eckstein2011,Averchenko2014}.
We then discus the results in Sec. \ref{sec:train-degauss} regarding the exact nature of the multimode quantum light, either a train of squeezed pulses \cite{Slusher1987,Wenger2004} or a train of correlated pulses \cite{Roslund2013}.
We conclude in Sec. \ref{sec:concl} by summing up our results.

\section{General framework of multimode single-photon subtraction}\L{general}

\subsection{The single-mode case}

In order to introduce the general formalism, we recall the results about the subtraction of a photon from a single-mode light field, whose density matrix writes $\hat{\rho}$.
A single-mode beamsplitter mixes the single-mode light field with vacuum.
We write their respective annihilation operators $\hat{A}$ and $\hat{B}$ along with their corresponding  optical modes $\vec{\a}(\vec{r},t)$ and $\vec{\b}(\vec{r},t)$.
We remind that a mode of light is simply defined as a normalized solution of Maxwell's equations \cite{Kolobov1999,Treps2005}.
Those optical modes contain every physical property of the light field such as space-time dependence, momentum, polarization, optical frequency, etc.
The single-mode beamsplitter is represented as a unitary transformation $\hat{U}$ and one can assume the interaction to be weak ($\tt\ll1$):

\begin{equation}
    \hat{U} = \exp \left[ i \tt ( \hat{B}^\dagger\hat{A} + \hat{B}\hat{A}^\dagger ) \right] \approx \hat{\mathds{1}} + i \tt ( \hat{B}^\dagger\hat{A} + \hat{B}\hat{A}^\dagger )
\end{equation}

On the split arm, a single-photon detector of unit quantum efficiency performs a single-photon detection with annihilation operator $\hat{D}$ on an optical mode denoted $\vec{\dd}(\vec{r},t)$.
The detection operator is a Positive Operator of Measurement (POM) \cite{Barnett2009} $\hat{\Pi} =  \hat{\mathds{1}} - \ket{0}\bra{0}$ that we intentionally reduce to a single-photon detection operator $\ket{1}\bra{1}_{\dd}$ in mode $\vec{\dd}(\vec{r},t)$.
The output signal conditioned on a single-photon detection can then be computed by performing a partial trace over the single-photon output subspace.
The conditioned density matrix readily reads:

    \begin{align}
    \hat{\rho}^- &= \text{Tr}_{|1\>} \left( \hat{U} (\hat{\rho} \otimes |0\>\<0| ) \hat{U}^\dagger \hat{\Pi} \right)
    / P \L{eq:single-mode-sub-dens-mat} \\
    &= \tt^2 |\<\vec{\b},\vec{\dd}\>|^2 \hat{A} \hat{\rho} \hat{A}^\dagger / P \notag\\
    \text{with } P &= \tt^2 |\<\vec{\b},\vec{\dd}\>|^2 \; \text{Tr} \left( \hat{A} \hat{\rho} \hat{A}^\dagger \right) \text{ and } \<\vec{\b},\vec{\dd}\> = \int \vec{\b}.\vec{\dd}^* d\vec{r} dt \notag
    \end{align}

Interestingly, $P$ is interpreted as a probability to subtract and detect a single photon and consequently depends on the number of photons of the input state and on the overlap $\<\vec{\b},\vec{\dd}\>$
\footnote{Exp. \eqref{eq:single-mode-sub-dens-mat} contains the factor $\bra{0}\hat{B}\ket{1}_\dd = \bra{0}\hat{B}\hat{D}^\dagger\ket{0}$ and its hermitian conjugate. In a modal description of the electro-magnetic field, it is proven to be equal to $\<\vec{\b},\vec{\dd}\>$
	}
 between the optical modes in which the single photon is carried and detected.

\subsection{A general framework for the multimode subtraction}

\begin{figure}[h]
	\center{\includegraphics[width=0.99\linewidth]{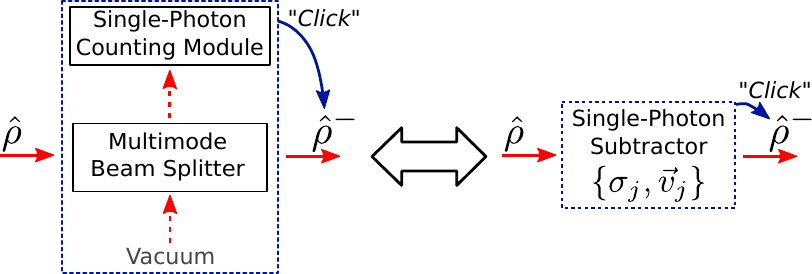}}
	\caption{Multimode subtraction of a single photon. Subtraction is modeled as the interaction of a multimode input with vacuum through a multimode beam splitter and the subsequent conditioning on the photon detection (left); this description is equivalent to a "single-photon subtractor" with subtraction eigenmodes $\{ \vec{v}_j \}$ and subtraction efficiencies $\{ \s_j \}$ (right)}
	\label{fig:scheme_combined}
\end{figure}

Here, we develop the general framework to describe the conditional single-photon subtraction from a quantum light in an arbitrary multimode state $\hat\rho$ based on the previous subsection.
The procedure is performed through the following steps: light splitting that can be performed via different physical mechanisms (linear beam-splitter, weak parametric up-conversion), filtering of the split light (spatially, spectrally) and conditioning the state of the signal light onto the detection of a single-photon in the split arm (see Fig.~\ref{fig:scheme_combined}). 

We choose to represent our multimode beam-splitter as a two-port device coupling some input light modes $\{\vec{\alpha}_n(\vec{r},t)\}$ of a signal port with some vacuum modes $\{\vec{\beta}_n(\vec{r},t)\}$ (whose we will drop the space-time dependence from now) of the optical bath.
Those two sets form two orthonormal basis of a function space
\footnote{This representation also covers cases where multiple input modes are mixed with multiple vacuum modes, regardless the number of modes involved and the basis chosen to express them. The overall transformation being unitary, a singular value decomposition will lead to the aforementioned representation.}
 with its scalar product $\<.,.\>$.
Their associated annihilation operators will be denoted $\hat{A}_n$ and $\hat{B}_n$.
The input state entering this multimode beam-splitter will be represented by the separable density matrix: $\hat\rho \otimes |0\>\<0|$.
The overall interaction between the two ports is summarized by the evolution operator $\hat{U}$:

	\begin{align}
	\hat{U} &= \exp \left[ i \sum_n \tt_n ( \hat{B}_n^\dagger\hat{A}_n + \hat{B}_n\hat{A}_n^\dagger ) \right] \nonumber \\
	&\approx \hat{\mathds{1}} + i \sum\limits_n \tt_n  ( \hat{B}_n^\dagger\hat{A}_n + \hat{B}_n\hat{A}_n^\dagger )
	\label{eq:evolution_op}
	\end{align}

We have assumed weak interaction ($\tt_n\ll1$) so that a first order Taylor development could be performed.
The two parties thus exchange a single photon at maximum.
Accordingly, we consider that a single photon is to be detected in the split arm.
Its detection operator $\hat{\Pi}$ can be represented as a mixture of single-photon measurements on a set of detection modes $\{\vec{ \dd}_m\}$ whose annihilation operators are denoted $\hat{D}_m$:
    \begin{align}
    & \hat{\Pi} =  \sum_m \g_m \ket{1_m} \bra{1_m}_{\dd} \label{eq:detection_op} \\
    \text{where:} \; & \ket{1_m}_{\dd} = |0,...,0,\underset{m}{1},0,...\>_{\dd}  = \hat D_m^\dag|0\> \notag \\
    	\text{and} \; & 0 \leq \g_m \leq 1 \notag
    \end{align}

Where the coefficients $\{\g_m\}$ are quantum efficiencies associated with the detection modes ${\dd_m}$ and represent losses.
The $\dd$ suffix meaning that the state is written in the basis of modes ${\dd_m}$.
One notes that the detection modes are a priori different from the beamsplitter modes.

The output signal conditioned on single photon detection can then be computed by performing a partial trace over the single photon output subspace
\footnote{The measurement being imperfect, the detection operator $\hat{\Pi}$ should eventually be written in terms of Kraus operators as:
    \begin{equation}
    \hat{\Pi} = \sum_i \hat{\kappa}_i \hat{\kappa}_i^\dagger \quad
    \text{with} \; \hat{\kappa}_i = \sqrt{\gamma_i} |1_i\> \<1_i|_{\dd} \notag
    \end{equation}
And the conditioned density matrix is computed as :
\begin{equation}
	\hat{\rho}^- = \text{Tr}_{|1\>} \left[ \sum_i \hat{\kappa}_i \hat{U} (\hat{\rho} \otimes |0\>\<0| ) \hat{U}^\dagger \hat{\kappa}_i^\dagger \right] / P \notag
\end{equation} 
    }
.
As a result the conditioned density matrix reads:

    \begin{align}
    \hat{\rho}^- &= \sum\limits_{n,n'} S_{nn'} \; \hat A_{n'} \hat \rho  \hat A_{n}^\dag/P \label{eq:sub_density_matrix} \\
    \text{with} \; & S_{nn'} =  r_n^* r_{n'} \sum_m \g_m \<\vec \b_{n},\vec d_m\>  \<\vec d_m, \vec \b_{n'}\> \label{eq:sub_coefs} \\
    \text{and} \; & P = \sum\limits_{n,n'} S_{nn'} \; \text{Tr}\(\hat A_{n'} \hat \rho  \hat A_{n}^\dag\) \label{P_def}
    \end{align}

The normalization constant $P$ defines the total probability to subtract a single photon for a given setup. The weak splitting condition, assumed during the derivation ensures $P \ll 1$.

Although distinct physical mechanisms can be applied to split light, exp. \eqref{eq:sub_density_matrix} shows that the subtraction of a single-photon is always characterized by a matrix $\mathbf{S}$ of coefficients $S_{nn'}$ that we will refer to as the subtraction matrix.
We know from exp. \eqref{eq:sub_coefs} that the matrix $\mathbf{S}$ is positive and hermitian.
It can then be eigen-decomposed with eigenvectors $\vec{v}_j$ and eigenvalues $\s_j$ so that exp. \eqref{eq:sub_density_matrix} becomes:

	\begin{align}
	& \hat{\rho}^- = \sum\limits_{j} \s_j \; \hat s_j \hat \rho  \hat s_{j}^\dag /P \label{eq:sub_density_matrix_eigen} \\
	\nn & \text{with:} \; \hat s_j = \sum\limits_{n} \<\vec v_j, \vec \a_n\> \hat A_n \notag
	\end{align}
	
Therefore the overall single-photon subtraction procedure involving multimode light splitting and detection, can be described in terms of orthogonal subtraction modes $\{\vec{v}_j\}$ with their annihilation operators $\hat{s}_j$ and efficiencies $\{\s_j\}$ (see Fig. \ref{fig:scheme_combined}). Interestingly, the efficiencies $\s_j$ are all smaller than unity according to exp. \eqref{eq:sub_coefs} and can be interpreted in terms of subtraction probability per subtraction mode as the total subtraction probability $P$ reads:

\begin{equation}
	P = \sum_j \; \s_j \text{Tr}\(\hat{s}_j^\dag \hat{s}_{j} \hat{\rho} \)
	\label{eq:P_exp}
\end{equation}

In general the single-photon subtraction is multimode and the conditioned state \eqref{eq:sub_density_matrix_eigen} is mixed. The efficient number of subtraction modes, which definition is similar to the one of a Schmidt number \cite{Ekert1995}, can be characterized with the quantity:

	\begin{equation}
	K = \frac{(\sum\limits_j \s_j)^2}{\sum\limits_j \s_j^2}
	\label{eq:Snum}
	\end{equation}

Expression \eqref{eq:sub_density_matrix_eigen} does not mean that there exist a basis where the conditioned density matrix $\hat{\rho}^-$ is diagonal.
Indeed, the annihilation operators $\hat{s}_j$ have no intrinsic reason to act on orthogonal subspaces.

Some insight into the properties of the conditioned state (\ref{eq:sub_density_matrix_eigen}) can be obtained by calculating its quantum marginals, i.e. reduced density matrices of its subsystems. Further we consider the properties of the conditioned state under different scenarios for the single-photon subtraction.

\section{Properties of the subtracted state}\L{sec:single_multi}

Expressions \eqref{eq:sub_density_matrix} and  \eqref{eq:sub_density_matrix_eigen} are general and applicable to describe single-photon subtraction from any arbitrary quantum state of light.
From now on, we consider a pure state of light such that there is a basis, defined by mode functions and associated bosonic operators $\{\vec{u}_k, \hat{a}_k\}$, in which the quantum state of the light is factorized:

	\begin{align}
	&\hat \rho = \bigotimes_{k\geq0} \hat{\rho}_{k}
	\label{eq:in_state}
	\end{align} 

Moreover, we consider that each single-mode state $\hat{\rho}_{k}$ is pure and has a null mean amplitude.
This assumption is valid for squeezed vacuum states, Fock states, cat states \cite{Ourjoumtsev2007} and their superpositions.
We introduce an estimator of the multimode character of the light field through the definition of an efficient number of non-vacuum modes $N$. This definition is inspired from the definition of the Schmidt number and is formally equivalent:
\begin{equation}
	N = \frac{(\sum n_k)^2}{\sum n_k^2}
	\label{Nnum}
\end{equation}
where $n_k$ is the mean photon number per mode $k$.
For further analysis, it is useful to introduce an expansion of the subtraction modes over the input modes:

	\begin{equation}
	\hat s_j = \sum\limits_j c_{jk} \hat a_k \quad \text{with} \quad c_{jk} = \<\vec{v}_j, \vec{u}_k\>
	\label{eq:sub_expansion}
	\end{equation}

\subsection{Matching the single-photon subtraction}
	
We will show that the photon subtracted state $\hat \rho^-$ is pure when the detected photon belongs with certainty to a particular mode.
It can result from different cases : the trivial case, when the light field is single mode \cite{Treps2005}, the interesting case, when the subtraction procedure is single-mode itself.
In the second case, there is only one term in the sum \eqref{eq:sub_density_matrix_eigen} and one can show that a unit purity is achieved with expression \eqref{eq:sub_expansion}.
Indeed denoting the corresponding annihilation operator of the subtraction with $\hat s$, the resulting pure state simply reads:
	\begin{equation}
	\hat \rho^- =  \hat{s} \; \hat \rho  \;\hat{s}^\dag /\text{Tr}\(\hat{s}^\dag \hat{s} \; \hat \rho \)
	\label{eq:pure_sub}
	\end{equation}
Nevertheless, while the state is pure, the photon subtraction does not necessarily happen in a eigenmode of \eqref{eq:in_state}.
Then the following scenarios can be realized:

\begin{itemize}
    \item In the case where $\hat{s}$ matches one of the $\hat{a}_k$ (matched case), the single-photon subtraction acts on a mode that we denote with index ''s'' so that :
	
	\begin{equation}
	\hat s = \hat a_{s} \nonumber \notag
	\end{equation}
	\begin{equation}
	\hat \rho^- = \hat \rho_{s}^- \bigotimes_{k \neq s} \hat\rho_k \quad \text{where} \quad \hat \rho_{s}^- = \hat a_s \; \hat \rho_s \; \hat a_s^\dag /n_s
	\label{eq:matched_sub}
	\end{equation}
	
and $n_s$ is an average number of photons in the mode. \\
\item In the general case (non-matched) the single photon is subtracted from a linear combination of the eigenmodes of \eqref{eq:in_state}:
	
	\begin{equation}
	\hat s = \sum \limits_k ~c_k ~\hat a_k \notag
	\end{equation}
	\begin{equation}
	\hat \rho^- \propto \( \sum\limits_{k_1,k_2}  
	c_{k_1} c_{k_2}^* \( \hat a_{k_1}\hat\rho_{k_1} \otimes \hat\rho_{k_2} \hat a_{k_2}^\dag \) \bigotimes_{k\neq k_1,k_2} \hat\rho_k \)
	\label{eq:mixed_sub}
	\end{equation}
	
A non-matched single-photon subtraction leads to entanglement of the light modes. It can be used to perform entanglement distillation \cite{Ourjoumtsev2007, Takahashi2010, Kurochkin2014} or to achieve a non-Gaussian gate on a node of an optically implemented cluster state \cite{Yokoyama2013, MedeirosdeAraujo2014}.
\end{itemize}

\subsection{Multimode state purity}

In general the photon subtraction is multimode and the output state $\hat \rho^-$ is mixed according to \eqref{eq:sub_density_matrix_eigen}. The state purity $\pi$ of the multimode light can be represented in two ways:
	\begin{align}
    	\pi = \mbox{Tr}((\hat\rho^-)^2) &=\sum\limits_{k,k'} n_k n_{k'} |\<\vec u_{k}, \mathbf{S} \vec u_{k'}\>|^2 /P^2 \L{eq:multimode_purity-1}\\
    	& = \sum\limits_{j,j'} \s_j \s_{j'} \left|\sum\limits_k c_{jk} c_{j'k}^* n_{k}\right|^2/P^2
	\label{eq:multimode_purity}
	\end{align}
The last expression provides a way to calculate the purity if the subtraction modes and associated efficiencies are known.
Also from \eqref{P_def} one gets expressions to calculate the photon subtraction probability:
	\begin{align}
    	P &= \sum\limits_{k} n_k \<\vec u_k, \mathbf{S} \vec u_{k}\> \L{P}\\
    	& = \sum\limits_{j,k} \s_j |c_{jk}|^2 n_{k} \L{P-1}
	\end{align}
One can distinguish two extreme cases for the output state depending on the relation between the numbers of squeezed modes $N$ and subtraction modes $K$.
In the case of non-selective photon subtraction, when $K \gg N$ (i.e. $\mathbf{S} \propto \hat{\mathds{1}}$) the state purity satisfies:

	\begin{equation}
    	\pi_{K \gg N}  = \frac{1}{N}
	\label{eq:purity_non-selective}
	\end{equation}

In the opposite case when $N \gg K$ (i.e. $n_k=\text{const}$), one can show that the purity reads:

	\begin{align}
    	\pi_{N \gg K} = \frac{1}{K}
	\label{eq:purity_multimode}
	\end{align}
	
The general statement is that the output state is pure when the detected photon belongs with certainty to a particular mode.
Either the light is single mode ($N=1$) \cite{Treps2005} or the single-photon subtraction procedure is single-mode itself ($K=1$).

\subsection{A two-modes example}

We illustrate the difference between pure and non-pure single-photon subtraction with a product state of two squeezed vacua:
\begin{equation}
	|\text{in}\> = |\text{sqz1}\> \otimes |\text{sqz2}\>
\end{equation}
In the situation where single-photon subtraction is not pure, the photon is subtracted with equal probabilities from each mode.
In the above formalism it means that there are two subtraction modes: $\hat{s}_{1,2} = \hat{a}_{1,2}$ and $p_1=p_2$. Then $c_{11}=c_{22}$ and the purity of the output state, according to (\ref{eq:multimode_purity}), becomes:

	\begin{equation}
			 \pi = \frac{n_1^2+n_2^2}{(n_1+n_2)^2} 
    \end{equation}
    
A near unit purity is then only achieved if one of the two modes totally overcomes the other with a much higher photon number i.e, if the input state is almost single mode.

When the single-photon subtraction is pure, the photon is subtracted coherently from a superposition of modes, for example: $\hat{s} = (\hat{a}_1 + \hat{a}_2)/\sqrt{2}$.
Then $c_{11} = c_{12} = 1/\sqrt{2}$ and the purity of the output state is then necessarily equal to unity.
The output state is a superposition of squeezed vacuum and single-photon subtracted squeezed vacuum:

	\begin{equation}
	\out \propto \bigg(\hat{a}_1|\text{sqz1}\rangle|\text{sqz2}\> + |\text{sqz1}\> \hat{a}_2 |\text{sqz2}\>\bigg)
	\end{equation}

As expected, the state is non longer factorizable in the original squeezing basis and the modes are entangled.
What is at stake is the interplay between the eigenmodes of the squeezed light and the single-photon subtraction modes.

\subsection{State of a single mode}

We consider the quantum state of an eigenmode ''s'' of the multimode squeezed state $\hat\rho$ after single-photon subtraction.
The conditional probability $p_s$ to subtract a single photon from the target mode among other field modes depends on the overlap of the mode with the subtraction modes and on the number of photons in all modes:
	\begin{align}	
	p_{s} &= n_s \<\vec u_s, \mathbf{S} \vec u_s\>/P \L{P0}\\
    	&= n_s\sum\limits_j \sigma_j |\<\vec u_s,\vec v_j\>|^2/P \L{P0-1}
	\end{align}
We denote the density matrix after multimode single-photon subtraction $\hat{\rho}^-|_s$.
It is calculated by tracing out the other modes in the multimode state $\hat\rho^-$ and can be generally written as a mixture of the single-photon subtracted single-mode state ($\hat{\rho}_s^-$ of eq. \eqref{eq:matched_sub}) and the initial state $\hat{\rho}_s$ of the mode:
	\begin{equation}
    	\hat\rho^-|_s = \text{Tr}_{k\neq s} \left( \hat{\rho}^- \right) = p_s \; \hat\rho_s^- + (1-p_s) \; \hat\rho_s
    	\L{eq:single-mode}
    	\end{equation}
As $\hat\rho_s$ is supposed to be pure, its fidelity with $\hat\rho_s^-$ is null and $p_s$ defines the fidelity \cite{Jozsa1994} of the resulting state $\hat{\rho}^-|_s$ with the single-photon subtracted state $\hat\rho_s^-$:
	\begin{align}
    	F(\hat{\rho}^-|_s, \hat\rho_s^-) = \text{Tr}(\hat{\rho}^-|_s \; \hat\rho_s^-) = p_s
	\L{F_single-mode}
	\end{align}
The purity of the state (\ref{eq:single-mode}) is calculated as:
	\begin{align}
    	\pi_s = \mbox{Tr}(\hat{\rho}^-|_s^2) = p_{s}^2 + (1-p_{s})^2 \leq 1
	\L{purity_single-mode}
	\end{align}
There are then two reasons for the single-mode state to be mixed.
Firstly, when the single-photon subtraction itself is not pure and results in a non-pure multimode state $\hat \rho^{-}$.
Secondly, if the subtraction mode is not matched to the targeted mode, (i.e. $|\<\vec v,\vec u_s\>|^2 < 1$) so that the single modes of the multimode state $\hat{\rho}$ become entangled and the state of a single mode is no longer pure.

\section{Photon subtraction from multimode squeezed vacuum}\L{sec:squeezed_light}

From now on, we consider the input signal light to be a pure multimode squeezed vacuum. The explicit expression for the state reads:
	\begin{equation}
	\hat \rho = \bigotimes\limits_k \exp\[ \frac{\xi_k}{2} \hat a_k^{\dag 2} - \frac{\xi_k^*}{2} \hat a_k^2 \] |0\> \; \<...|
	\label{eq:in_state_squeezed}
	\end{equation} 
where $\xi_k = r_k e^{i \theta_k}$  is a complex squeezing parameter of k-th mode.

\subsection{Negativity of the Wigner function}

We use exp. \eqref{eq:single-mode} to explicit the Wigner function of the state of a given eigenmode "s"of the multimode input state \eqref{eq:in_state_squeezed} :
	\begin{align}
    	W^-|_s(\a,\a^*) = p_s W_s^-(\a,\a^*) + (1-p_s) W_s(\a,\a^*)
	\L{W_single-mode}
	\end{align}
where $W_s,W_s^-$ are respectively the Wigner functions of the squeezed vacuum state and the single-photon subtracted squeezed vacuum state \cite{Biswas2007}, also known as a ``\textit{squeezed} single-photon state'':
	\begin{align}
    	W_s(\a,\a^*) &= \frac{2}{\pi} e^{-2|\tilde \a|^2},\\
    	W_s^-(\a,\a^*) &= \frac{2}{\pi} (4|\tilde \a|^2-1)e^{-2|\tilde \a|^2}
    	\L{W_single-mode-1}
	\end{align}
where $\tilde\a = \a \cosh r_s - \a^* e^{i \theta_s} \sinh r_s$ is a squeeze coordinate transformation.
Both the Wigner functions have extrema at the origin of the phase-space: $W_s(0,0)=-W_s^-(0,0)=2/\pi$.
In turn, the total Wigner function (\ref{W_single-mode}) of the state embedded in mode ''s'' possesses a negative value at the origin of the phase space only when $p_s > 1/2$.
This value constitutes a benchmark for the single-mode subtraction probability.

\subsection{Photon subtraction from weakly squeezed multimode light}\L{weakly-squeezed}

In this section, we consider the single-photon subtraction from weakly squeezed multimode light.
One might be willing to maximize the fidelity of the heralded state with a single-photon state.
A weakly squeezed state is a superposition of mostly vacuum and a pair of photons\footnote{To illustrate this concept, one can consider a single mode where the probability to measure a two-photon state must be much greater compared to the probability to measure a four-photon (or higher even number) state. A ratio of $10$ corresponds to about $3.17dB$ of squeezing.} delocalized in all modes.
Under this assumption, the state (\ref{eq:in_state_squeezed}) can be approximated as:
	\begin{align}
	\hat \rho \approx \(|0\> + \frac{1}{\sqrt{2}}\sum\limits_k \xi_k |2_k\>_\text{u} \) \; \<...|
	\label{eq:weak_sqz}
	\end{align}
where we defined Fock states of the eigenmodes {$\vec{u}_k$}: $|n_k\>_\text{u} = |0,...,0,\underset{k}{n},0,...\>_\text{u} = (\hat a_k^\dag)^n|0\> / \sqrt{n!}$.
Under this approximation one can get explicit results for the single-photon subtracted state that can be treated as an approximation of the general case for arbitrary squeezing. 

The conditional subtraction of a photon heralds a single-photon state. The corresponding state is derived from exp. (\ref{eq:sub_density_matrix_eigen},\ref{eq:sub_expansion},\ref{eq:weak_sqz}) and can be written as:
	\begin{align}
	\hat \rho^- & = \sum\limits_{k,k'} L_{kk'} \;|1_{k'}\>\<1_{k}|_\text{u} / P
	\L{single_ph_nondiag}
	\end{align}
with the coefficients:
	\begin{align}
	L_{kk'} & = \xi_k^* \xi_{k'} \<\vec u_{k},\textbf{S} \vec u_{k'}\> \notag\\
	& = \xi_k^* \xi_{k'} \sum\limits_j \s_j  \<\vec u_{k},\vec v_j\> \<\vec v_j, \vec u_{k'}\> \L{Mkk}
	\end{align}
We first consider the case where the single-photon subtraction is single-mode and happens in mode $\vec v$. It is easy to show that instead of exp. \eqref{single_ph_nondiag}, ones simply gets a pure single-photon state:
	\begin{align}
	& |1\> \propto \sum\limits_k \xi_k \<\vec v, \vec u_k\> \; |1_k\>_\text{u}
	\end{align}
The single photon is then heralded in a mode $\vec w$ that can be written in the basis where $\hat{\rho}$ as a superposition (non-normalized) of modes: $\vec w \propto \sum\limits_k \xi_k \<\vec v, \vec u_k\> \; \vec u_k$.
This superposition is thus defined by the squeezing parameter of each input mode as well as its overlap with the subtraction mode.
\footnote{As an illustration, consider photon subtraction from a CW squeezed light from a degenerate optical parametric oscillator operating below the parametric threshold \cite{Neergaard-Nielsen2007}. The squeezed light possesses sideband squeezing in the bandwidth of resonance of the oscillator cavity. One can model it as frequency dependent squeezing $\xi_k \rightarrow |\xi_\w| \propto (1+4\w^2/\g^2)^{-1}$ of continuous modes of squeezing $\vec u_k \rightarrow u_\w(t) \propto \text{e}^{\text{i}\w t}$. Here we denote FWHM of the cavity resonance with $\g/2\pi$.
Consider instant subtraction of a single-photon from this light. One can model it with the following delta-like subtraction mode $\vec v \rightarrow \dd(t)$. Then one gets for the mode of the heralded photon $\vec w \rightarrow \int \ud\w \text{e}^{\text{i}\w t}/(1+4\w^2/\g^2) \propto \text{e}^{-\g |t|}$.
It is a double-sided exponent reproduced in different works on the photon subtraction \cite{Molmer2006, Nielsen2007, Neergaard-Nielsen2007, Morin2013}. We have illustrated its derivation using the developed multimode approach.}

Conversely, when the subtraction is not-selective at all (i.e. $\mathbf{S} \propto \mathbf{I}$ in exp. (\ref{Mkk})), the heralded single-photon state of exp. (\ref{single_ph_nondiag}) reads: $\hat\rho^{-} = \sum |\xi_k|^2 \; |1_k\>\<1_k|_\text{u} / P$.
The single-photon is then heralded with the highest probability in the most squeezed mode.
	
To treat the intermediate situation, we introduce an hermitian matrix $\mathbf{L}$ of the single-photon state similarly to the subtraction matrix $\mathbf{S}$.
This matrix can be eigen-decomposed on a basis with a set of orthogonal eigenvectors $\{\vec{w_l}\}$ so that:
	\begin{equation}
	\textbf{L} = \sum\limits_{k,k'} L_{kk'} \, \vec u_{k} \,  \vec u_{k'}^\dag = \sum\limits_l \l_l 	\, \vec w_l \,  \vec w_l^\dag
	\label{eq:L_eigen}
	\end{equation}

The substitution of exp. (\ref{eq:L_eigen}) into exp. (\ref{single_ph_nondiag}) gives the single-photon density matrix in a diagonal form, i.e. as a sum of single-photon states in orthogonal modes:

	\begin{align}	
	 \hat \rho^- &= \sum\limits_{l} \l_l \;|1_l\>\<1_{l}|_\text{w} / P, \label{eq:fock}\\
	& \text{where:} \; |1_l\>_\text{w} \propto \sum\limits_k \<\vec w_l,\vec u_k\> |1_k\>_\text{u} \notag
	\end{align}

The highest eigen-value in the decomposition defines the mode where the photon is heralded with the highest probability. In general, this mode is different from the eigenmodes of the original input state. We denote this mode with the index ''f'', the state embedded in this mode is a statistical mixture of a single-photon state and vacuum state:
	
	\begin{align}
	 \hat {\rho}_f &= \text{Tr}_{l\neq f} \left( \hat{\rho}^- \right) = p_f \; |1\>\<1|_f + (1-p_f) |0\>\<0|_f, \label{eq:rho_fock}\\
	& \text{where:}\; p_f = \l_f/P \notag
	\end{align}
	
The coefficient $p_f$ gives the conditional probability to herald a single photon in the mode ''f''.
It is also the fidelity of the state in mode ''f'' with the single-photon Fock state.
The Wigner function of the state in mode ''f'' can be written with the expressions (\ref{W_single-mode},\ref{W_single-mode-1}), where one have to set $\tilde \a = \a$.
It leads to a statistical mixture of vacuum and a single-photon state.
Then, the Wigner function has negativity at the origin of the phase space only when $p_f >1/2$. Furthermore, since the following relation always holds: $p_f \geq p_s$
\footnote{
Comparing \eqref{P0} and \eqref{Mkk} alongside with \eqref{eq:rho_fock} one sees that $p_s$ is a diagonal element of the matrix $L$, while $p_f = \text{eigenvalue}_\text{max}(L)/P$ is the maximal eigenvalue of the matrix. 
Then the relation holds: $p_f \geq p_s$ according to the Courant-Fischer theorem.
}
, the Wigner function of mode "f"  will be the better candidate to find negativity.

To summarize, we have calculated the purity of the multimode squeezed light after the subtraction of a single photon and have derived the conditions to get a pure output multimode state.
We have calculated the reduced density matrices in different light modes.
Firstly, we have calculated the state of an individual mode of squeezing $\hat\rho^-|_s$ completely defined by the conditional probability $p_s$ of the single-photon subtraction from this very mode.
In turn, when the light is weakly squeezed the photon subtraction heralds the light in a single-photon state.
We have then found a mode where the single-photon state is heralded with the highest conditional probability $p_s$.

\section{Photon subtraction from spectrally/temporally multimode light}\L{sec:time-domain}

The results obtained above are applicable to describe the photon subtraction from any type of light modes.
Herein, we focus on spectrally/temporally multimode light and consider two subtraction methods performed via traditional beamsplitting \cite{Dakna1996, Wenger2004} and via weak parametric up-conversion, recently studied in \cite{Averchenko2014}.
In the second case the subtraction of a photon from signal light is efficiently performed via its parametric up-conversion in a non-linear medium by strong gate field. 

\subsection{Linear/non-linear subtraction}\L{L-NL BS}

Both linear and non-linear methods of the photon subtraction are of probabilistic nature.
We assume that the probability to extract more than one photon from a signal field is negligibly small. It is achieved using a low reflectivity beamsplitter or weak up-conversion. 
The extraction of a photon from signal a field with annihilation operator $\hat a$ into an auxiliary field with creation operator $\hat b^\dagger$ can be described in the frequency domain by the following operator:

	\begin{align}
	\hat U \approx \mathds{1} + i \iint \ud\w \ud\w' \; R(\w,\w') \, \hat b^\dag(\w) \, \hat a(\w')
	\L{out_AB}
	\end{align}

The expression is the counterpart of the exp. (\ref{eq:evolution_op}) in the frequency domain.

In the case of a beamsplitter the interaction kernel is diagonal in the frequency domain meaning that photon energy is preserved during the exchange:

	\begin{align}
	R(\w,\w') = r \;\dd(\w-\w') \quad \text{(beamsplitter)}
	\L{state_BS}
	\end{align}

With $r$ being the amplitude reflection coefficient, such that $r\ll1$.

In the up-conversion, input and converted photons have different energies compensated by the energy of the gate field.
The spectrum of the gate field is described by the normalized function $\a(\w)$, such that $\int\ud\w |\a(\w)|^2=1$. The process efficiency is governed by the phase-matching condition in the non-linear medium described by the function $\Phi(\w,\w')$. While the up-conversion can be performed in different geometries/regimes (e.g. non-collinear \cite{Averchenko2014}, collinear \cite{Eckstein2011}), we consider a model case described by the following interaction kernel:
	\begin{align}
	\nn R(\w,\w') &= C \; \a(\w-\w') \Phi(\w,\w')\\
	&=C\sum\limits_{n=0}^\8 r_n \psi_n(\w) \varphi_n^*(\w') \quad \text{(up-conversion)}
	\L{Lup-conversion}
	\end{align}
Expression \eqref{Lup-conversion} is obtained through Mercer's theorem and represents the Schmidt decomposition of the kernel of parametric interaction \cite{Law2000}.
The decomposition shows that the parametric process operates as a beamsplitter for broadband modes \cite{Eckstein2011, Brecht2014} at signal and up-converted carrying frequencies $\{\varphi_n, \psi_n\}$, respectively, with the reflection coefficients $\{r_n\}$ and multiplier $C$ which is proportional to the length of the non-linear medium and square root of the energy of the gate pulses \cite{Averchenko2014}.

Spectral filtering of an extracted photon with the transmission $F(\w)$ (such that $|F(\w)|\leq1$) can be described in the expression \eqref{out_AB} via the replacement:
	\begin{align}
	& \hat b^\dag(\w) \rightarrow F(\w) \hat b^\dag(\w)
	\L{filtering}
	\end{align}
The expression requires two comments. Firstly, we assume post-selection on the successful detection photons that passed the filter.
Therefore we have omitted the term that describes loss of the photon due to the filtering.
Secondly, the filter function $F(\w)$ enters into the expression without complex conjugation as one could expect.
It reflects the following time ordering of physical processes: firstly, beamsplitting according to \eqref{out_AB} and, secondly, subsequent filtering. Conjugation of the function would correspond to the reversed order of these processes.

\subsection{Time-resolved detection of a photon}

In this section we refine the definition of the single-photon POM $\hat\Pi$ with particular emphasis on time-resolving measurement. Let us note that the modeling of frequency resolving photon detection has been considered in a number of papers, such as  \cite{Rohde2006,Tualle-Brouri2009}.

The photon detection can be performed with a detector that does not resolve number of photon or with a photon resolving detector.
A non photon-number resolving detection conditions the signal field into a mixed state in general \cite{Barnett1998}.
We will assume that the probability to have more than one photon in the split field is negligibly small.
This is guaranteed by the assumption that the splitting is weak.
Therefore click for both types of detectors can be treated as a detection of a single photon.

Time-resolved measurement of a single photon can be modeled as the projection onto an "instant" single-photon state: 
	\begin{align}
	\hat\Pi_t &= |1_t\> \<1_t| \\
	\text{with} \quad |1_t\> = \hat b^\+(t)|0\> &= \(\int \hat b(\w) e^{-i\w t} \ud\w/\sqrt{2\pi}\)^\+|0\> \notag
	\end{align}

Realistic photodetectors have a finite temporal resolution (due to temporal response of the electronics, known as jitter) that makes the click appearance classically random in some time interval.
It becomes particularly important in the detection of short optical pulses.
We model it through a detector response function $\gamma(t)$, that defines a time window of the detector response if a photon hits the detector at time instant zero.
Assuming stationary detector properties, we describe the detection operator at the moment $t$ as a statistical mixture of "instant" photodetections:
	\begin{align}
	\hat{\overline{\Pi}}_t = \int \ud \tau \; \g(\tau-t) \, \hat\Pi_\tau
	\L{P_t}
	\end{align}
	
In the limiting case of time non-resolving photodetection, i.e. $\g(t)=1$, the operator takes the following form:
	\begin{equation}
	\hat{\overline{\Pi}}= \int \ud t \; |1_t\> \<1_t| = \int \ud\w \; |1_\w\>\<1_{\w}|
	\L{Paverage}
	\end{equation}
where we also introduced a monochromatic single-photon state: $|1_\w\>=\hat b^\dag(\w)|0\>$. Then last equality means that detected photon is also not resolved in
frequency.

\subsection{Conditioned density matrix}

Detection of a photon in a split beam results in a conditionally single-photon subtracted state of the signal light. The state is calculated as follows:

	\begin{align}
\hat{\rho}^- \propto \text{Tr}_B \left[ \hat{U} (\hat{\rho} \otimes |0\>\<0|_B ) \hat{U}^\dagger \hat{\Pi} \right]
 	\end{align}

using the evolution operator \eqref{out_AB}, the measurement operator \eqref{Paverage} and tracing out the resulting density matrix over the basis of the split beam. 
Here we consider the case of slow detector as this detection regime is of particular importance to describe the single-photon subtraction from the pulsed quantum light that we will consider in the next section.
The general case is treated in Appendix \ref{Kgeneral}.
Then the conditioned density matrix reads:
	\begin{align}
	& \hat\rho^- \propto  \iint \ud\w \ud\w' \; S(\w,\w') \; \hat a(\w') \hat\rho \hat a^\dag(\w)
	\end{align}

The subtraction kernel for a beamsplitter is diagonal in the frequency domain:
	\begin{align}
	S(\w,\w') = r^2 \; |F(\w)|^2 \dd (\w-\w') \;\; \text{(beamsplitter)}
	\L{K_BS_slow}
	\end{align}
For weak up-conversion we assume no filtering of up-converted photons (for general case see Appendix \ref{Kgeneral}) and get:
	\begin{align}
	& S(\w,\w') = |C|^2 \sum\limits_j r_j^2 \varphi_j(\w) \varphi_j^*(\w') \quad \text{(up-conversion)}
	\L{K_QPG_slow}
	\end{align}
Expression (\ref{K_QPG_slow}) means that the subtraction modes are defined by the modes of the up-conversion process and the subtraction efficiencies are given by the squares of the up-conversion efficiencies (i.e. $v_j(\w) = \varphi_j(\w)$ and $\s_j = |C|^2 r_j^2$).

In the following section we compare these two methods for photon subtraction from particular types of multimode light.

\section{Degaussification of pulsed multimode squeezed light}\L{sec:train-degauss}

In this section we consider two types of temporally/spectrally multimode quantum light produced by different experimental setups.
Firstly, we consider an individual short pulse of squeezed light generated by single-pass degenerate parametric down conversion of a pump pulse in a non-linear medium such as a bulk crystal \cite{Slusher1987, Wenger2004} or a non-linear wave-guide \cite{Avenhaus2008, Eckstein2011}. Secondly, we consider a train of correlated pulses, known as a frequency comb \cite{Roslund2013} hosting a multimode squeezed quantum state.

\subsection{Degaussification of squeezed single pulses}
\L{pulses}

A single pulse can be represented in general as a tensor of broadband independently squeezed modes \cite{Wasilewski2006,Christ2011}. The modes can be approximated by Hermite-Gaussian functions for the Gaussian pump pulses driving the parametric down-conversion.
Using the notation of the Sec. \ref{sec:single_multi} one can write for the eigenmodes in the frequency domain:
	\begin{align}
	& \vec u_k \rightarrow u_k(\w) 
	\propto \text{H}_k(\t \w) e^{-\t^2 \w^2/2}
	\L{HG_modes}
	\end{align}
where H$_k$ is the Hermite polynomial of the k-th order. We assume that there are $N$ squeezed modes with $n_k$ photons per mode. Also odd and even modes are squeezed in orthogonal quadratures \cite{Patera2009}. To summarize:
	\begin{align}
	\xi_k \approx  (-1)^{k}\sqrt{n_k}, \quad (k = 0, 1, 2, \ldots)
	\L{N_modes}
	\end{align}
within the approximation of weak squeezing.

\begin{figure}[b]
	\center{\includegraphics[width=0.9\linewidth]{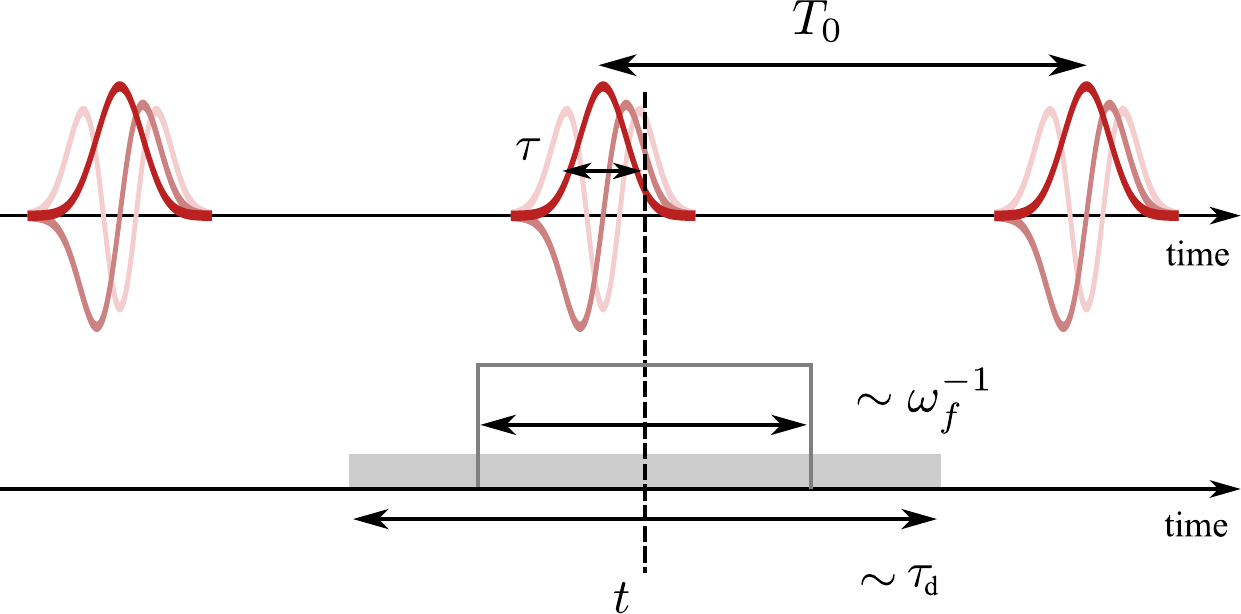}}
	\caption{Timing of the photon subtraction. (Top) train of multimode squeezed pulses (for details see Sec.\ref{sec:train-degauss}); (Bottom) click of the detector at the time instant $t$ heralds subtraction of a photon within a coherence time window defined by the photon filtering bandwidth $\w_f^{-1}$ (rectangular box represents characteristic time window). Photodetector jitter leads to classical uncertainty of the click instant within the interval $\t_d$ (marked by the gray interval).}
	\label{fig:subtraction_timing}
	\end{figure}

Usually in the experiment a train of independently squeezed pulses is generated using a periodic pump \cite{Wenger2004}.
Fig. \ref{fig:subtraction_timing} illustrates timing of the photon subtraction at the time instant $t$ qualitatively preserving a scale between main parameters: duration of pulses $\t\;\text{[fs-ps]} <$ coherence time of filtered photons $\w_f^{-1}\;\text{[fs-ps]} <$ jitter of a photodetector $ \t_d\;\text{[100 ps]}<$ period of pulses $T_0\;\text{[10-1000 ns]}$. 
Then the photon subtraction is selective with respect to individual pulses.
On the other hand the detector performs averaging over a pulse and can be described by the expression (\ref{Paverage}) as infinitely slow on the time scale of a pulse duration.

\subsubsection{Photon subtraction with beam-splitting and filtering}

\begin{figure}[h]
	\center{\includegraphics[width=0.99\linewidth]{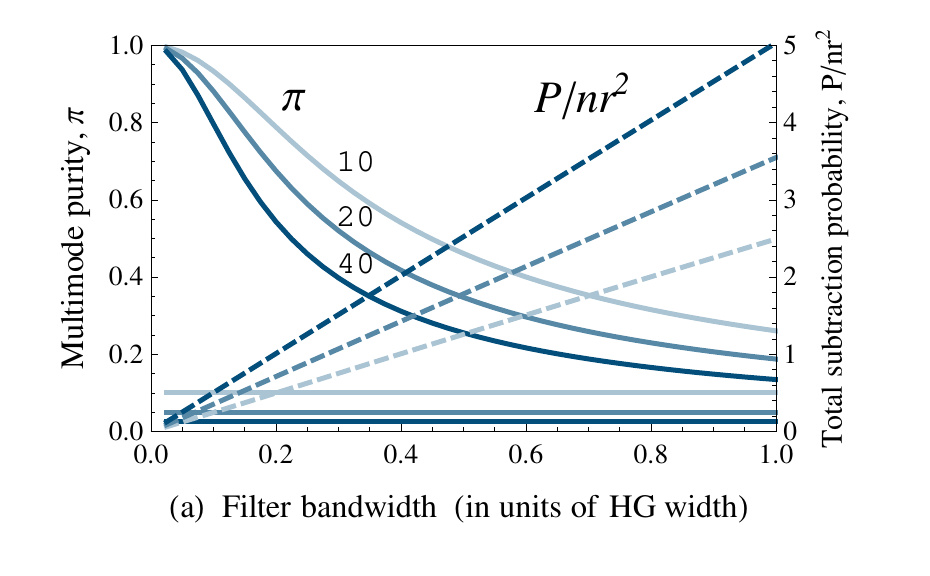}}
	\caption{Purity of a single-photon subtracted multimode state (solid lines) and its total subtraction probability (dashed lines) with respect to the Gaussian spectral filter bandwidth for $N=10, 20, 40$ equally squeezed Hermite-Gaussian modes. Total subtraction probability is normalized to the number of photons per mode $n$ and to the reflection coefficient of a beamsplitter $r$. Horizontal lines give asymptotic purities without filtering.}
	\label{fig:BS_results_multi}
\end{figure}

The photon subtraction probability from a single pulse and the purity of the resulting multimode state are obtained substituting the corresponding kernel (\ref{K_BS_slow}) into expressions \eqref{eq:multimode_purity-1} and \eqref{P} and taking into account identity \eqref{HG_modes} and state property \eqref{N_modes}. 
One gets for the purity :
	\begin{align}
	& \pi= \frac{\sum\limits_{k,k'=0}^{N-1} n_k n_{k'} |S_{kk'}|^2 }{P^2}
	\end{align}
and the subtraction probability :
	\begin{equation}
	P =  \sum\limits_{k=0}^{N-1} n_k S_{kk}
	\L{purity_BS_slow}
	\end{equation}
where elements of the subtraction matrix in the squeezing basis are defined
	\begin{align}
	S_{kk'} = \<\vec u_{k}, \textbf{S} \vec u_{k'}\> = r^2 \int \ud\w \; |F(\w)|^2 u_{k}^*(\w) u_{k'}(\w)
	\L{d-kk}
	\end{align}
These elements are proportional to the overlap between the filtered eigenmodes of the input multimode state.
To get explicit results we model the spectral transmission of the filter with a Gaussian function of bandwidth $\w_f$ centered at the carrier frequency of the signal light:
	\begin{align}
	& |F(\w)| = e^{-\w^2/2\w_f^2}
	\end{align}
One notes that the overlap matrix in this case can be calculated analytically
\footnote{
From \cite[eq. 7.374-5]{Gradsshteyn2007} it follows:
$\nn S_{kk'} \propto 2^{\frac{k+k'-1}{2}}\a^{-k-k'-1}	(1-2\a^2)^{\frac{k+k'}{2}} \G\(\frac{k+k'+1}{2}\) {}_2F_1(-k,k';\frac{1-k-k'}{2};\frac{\a^2}{2\a^2-1})$ when $k+k'$ is even, otherwise zero, and $\a^2=\frac{1}{2}\(\frac{1}{\w_f^2 \tau^2}+1\)$.
}.
Such a symmetric filter cancels odd Hermite-Gaussian modes because of their parity. It thus increases the selectivity of the photon subtraction and enhances purity of the resulting state. Fig.~\ref{fig:BS_results_multi} represents multimode state purity alongside with the normalized subtraction probability $P/n r^2$ (normalizing by $nr^2$ allows us not to choose any specific values for $n$ and $r$). Both quantities depend on the filter bandwidth and on the number of squeezed modes. In the limit of broad filtering the subtraction is non-selective and purity on the right hand side tends to the limiting values given by inversed number of eigen-modes (see (\ref{eq:purity_non-selective})). On the other side strong filtering results in the pure photon subtracted multimode state at the expense of reduced subtraction probability. More detailed consideration in the Appendix \ref{app:Gauss-decomp} shows that the resulting state is pure when the filter bandwidth is narrower than inversed temporal resolution of the detector.

\begin{figure}[h]
	\includegraphics[width=0.99\linewidth]{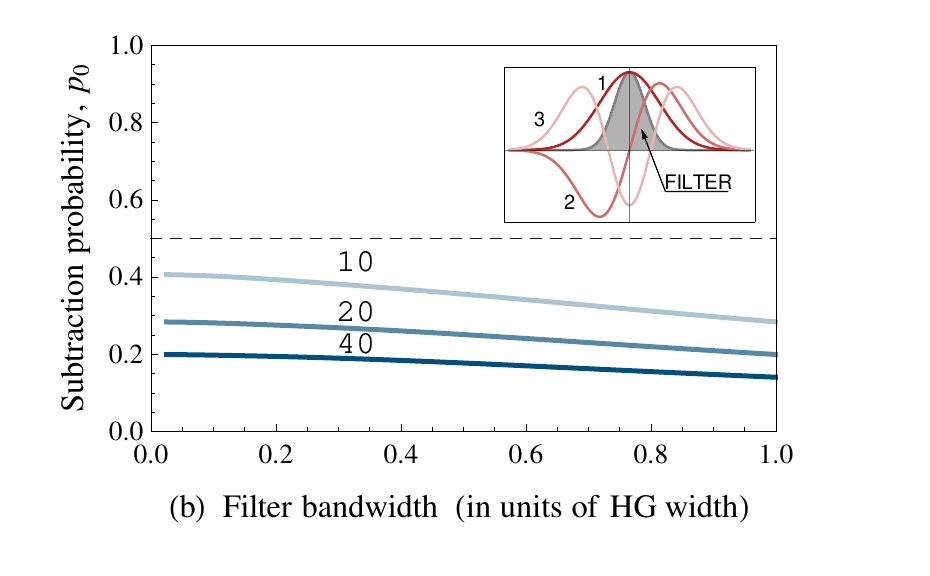}
	\caption{Conditional subtraction probability from the first Gaussian mode of the squeezing (solid lines) with respect to the Gaussian spectral filter bandwidth for $N=10, 20, 40$ equally squeezed Hermite-Gaussian modes. Inset depicts the profiles of the first three squeezed modes alongside with Gaussian transparency window of a spectral filter. Dashed horizontal line marks the threshold above which the corresponding Wigner function has negative values}
	\label{fig:BS_results_single}
\end{figure}

Let us characterize a state of an individual mode of squeezing that we denote with index ''s''. As it is shown in (\ref{eq:single-mode}) the state is defined by the conditional probability $p_s$ of the photon subtraction from the mode. Using (\ref{P0}) one gets   
	\begin{align}
	p_s= \frac{n_s S_{ss}}{\sum\limits_{k=0}^{N-1} n_k S_{kk}}
	\L{eq:purity_single_BS_slow}
	\end{align}
Then Fig.~\ref{fig:BS_results_single} represents the subtraction probability calculated for the first Gaussian squeezed mode.
It is also the mode that possesses the highest value of the subtraction probability in comparison with the higher order modes.
For the considered parameters the probability $p_s$ is inferior to the critical value of $1/2$. Thus the Wigner function \eqref{W_single-mode} of the state of this particular mode does not possesses negativity.

\begin{figure}[h]
	\includegraphics[width=0.99\linewidth]{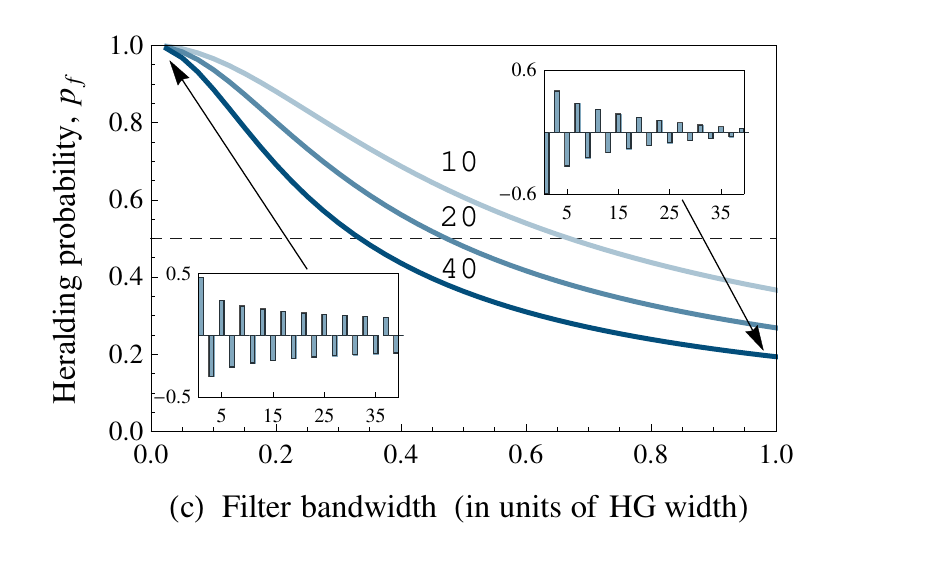}
	\caption{Heralding probability of a single-photon state for weakly squeezed multimode light with respect to the Gaussian spectral filter bandwidth for $N=10, 20, 40$ equally squeezed Hermite-Gaussian modes. Insets depict the mode that embeds the state with highest fidelity to the single-photon state in the basis of input squeezed modes.}
	\label{fig:BS_results_fock}
\end{figure}

Let us consider in more details the case of weakly squeezed light. The above results do not depend on squeezing parameter and are valid. On the other hand the photon subtraction heralds the light in the single photon state. One can find a mode, that we denote with index ''f'', where the photon is heralded with the highest probability. As it is shown in Section \ref{weakly-squeezed} it is the eigenmode of the matrix \eqref{Mkk} with the largest eigen-value. For considered in this section conditions \eqref{N_modes} and \eqref{d-kk} the matrix reads: $L_{kk'} \propto (-1)^{k+k'} \; S_{kk'} = S_{kk'}$, where $k,k'=0 \ldots N-1$.
The last equality holds, since $S_{kk'}$ is zero for odd $k+k'$. Then the largest heralding probability and corresponding single-photon mode can be calculated as follows
	\begin{align}
	& p_f= \frac{ n_f \quad \text{eigenvalue}_\text{max}(\textbf{S})}{\sum\limits_{k=0}^{N-1} n_k S_{kk}},\\
	& {\vec w}_f \propto \text{eigenvector}_\text{max}(\textbf{S})
	\end{align}
Fig.~\ref{fig:BS_results_fock} represents for different filtering both the conditional probability of a single photon and the corresponding mode in the basis of squeezed modes (see insets).
For narrow filter on the left hand side the subtraction heralds with unit probability a pure single-photon state.
This narrowband photon is distributed between modes of squeezing according to their amplitudes at the origin.
For moderate filtering the probability decreases, but still above one-half threshold for experimentally achievable region.
In this region the corresponding Wigner function possesses negative values. 
In the no-filtering limit the subtraction becomes non-selective. The photon is heralded in each equally squeezed mode with the probability $1/N$.

\subsubsection{Photon subtraction via weak up-conversion}

\begin{figure}[h]
	\center{\includegraphics[width=0.99\linewidth]{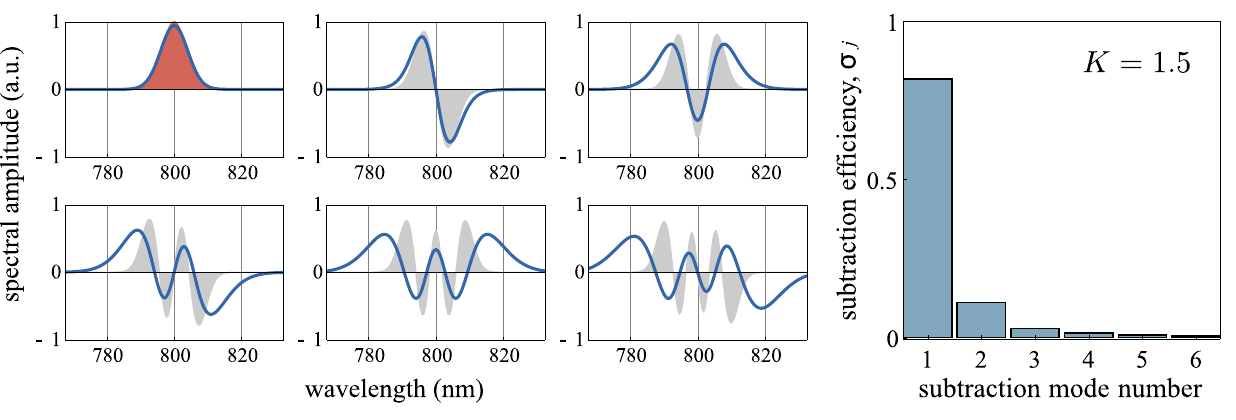}}
	\caption{(Left) Spectral profiles of subtraction modes of weak parametric up-conversion (solid blue) and eigen-modes of squeezing (filled grey); (Right) Normalized distribution of subtraction efficiencies. Inset denotes efficient number of subtraction modes (Schmidt number). Parametric up-conversion is gated with Gaussian pulses. For the details of the up-conversion see \cite{Averchenko2014}.}
	\label{fig:QPG_modes}
	\end{figure}

We consider the photon subtraction from a multimode squeezed pulse by weak parametric interaction with a synchronized gate pulse in a non-linear medium.
Such a non-collinear configuration is described in details in \cite{Averchenko2014}.
Fig.\ref{fig:QPG_modes} illustrates the corresponding spectral profiles $\{v_j(\w)\}$ of the first six subtraction modes estimated for a gate pulse spectrum matched to the first Hermite-Gaussian mode leading to a Schmidt number of about $1.5$.
The Hermite-Gaussian profiles $\{u_k(\w)\}$ of modes of squeezing are also represented along with the distribution of subtraction efficiencies $\{\s_j\}$ normalized to unity.

\begin{figure}[h]
	\includegraphics[width=0.99\linewidth]{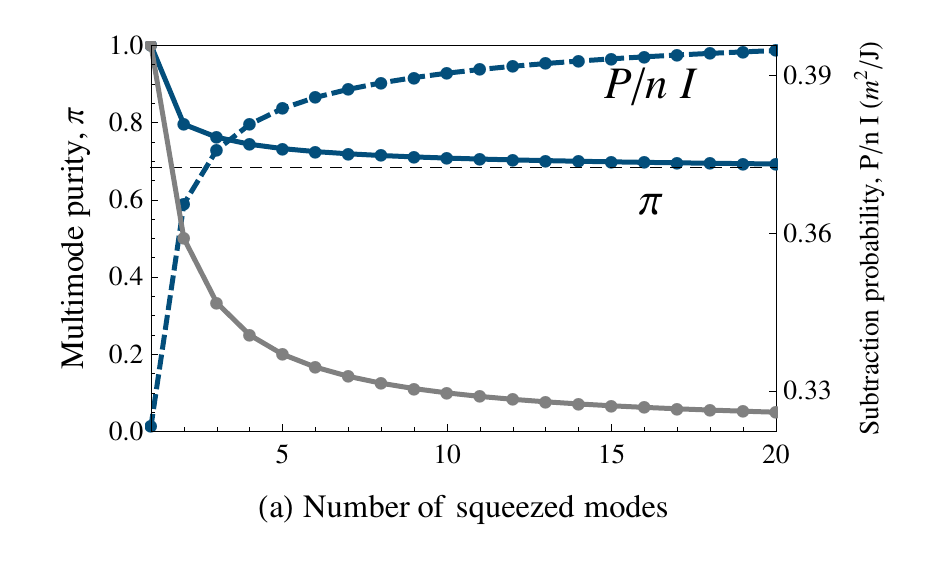}
	\caption{Purity of multimode state (solid blue line) after single-photon subtraction performed via weak parametric up-conversion with respect to the number of equally squeezed Hermite-Gaussian squeezed modes and with parameters depicted at Fig.\ref{fig:QPG_modes} and non-selective (solid grey line). Total subtraction probability (dashed blue line) is normalized to the number of photons per mode $n$ and average energy of gate pulses per unit area $I$}
	\label{fig:purityQPG_multi}
	\end{figure}

We calculate the overlap coefficients between subtraction modes and modes of squeezing, i.e. $c_{jk} = \int\ud\w \; v_j^*(\w) \;  u_k(\w)$.
We also assume that there are $N$ equally squeezed modes in the signal light possessing equal number of photons, i.e. $n_{k=0\ldots N-1}=n$.
It is then straightforward to calculate the purity of the heralded multimode state \eqref{eq:multimode_purity}.
To calculate the subtraction probability (\ref{P-1}), one needs to know the absolute values of $\s_j$ computed in \cite{Averchenko2014}.
The results are shown on Fig.~\ref{fig:purityQPG_multi} with respect to the number of equally squeezed modes.
Starting from a pure state on the left hand side the state purity decreases for a multimode light and reaches a limiting value given by the inversed number of subtraction modes $K^{-1}$.
One notices that the non-selective photon subtraction, depicted by the grey curve, results in a noticeably smaller state purity.
The depicted subtraction probability $P/n I$ is normalized to the number of photons per mode and to average energy of gate pulses per unit area.
One can check with expression (\ref{Lup-conversion})) that this probability increases slightly with the number of squeezed modes, meaning that photons are mostly subtracted from the first squeezed modes.

\begin{figure}[h]
	\includegraphics[width=0.99\linewidth]{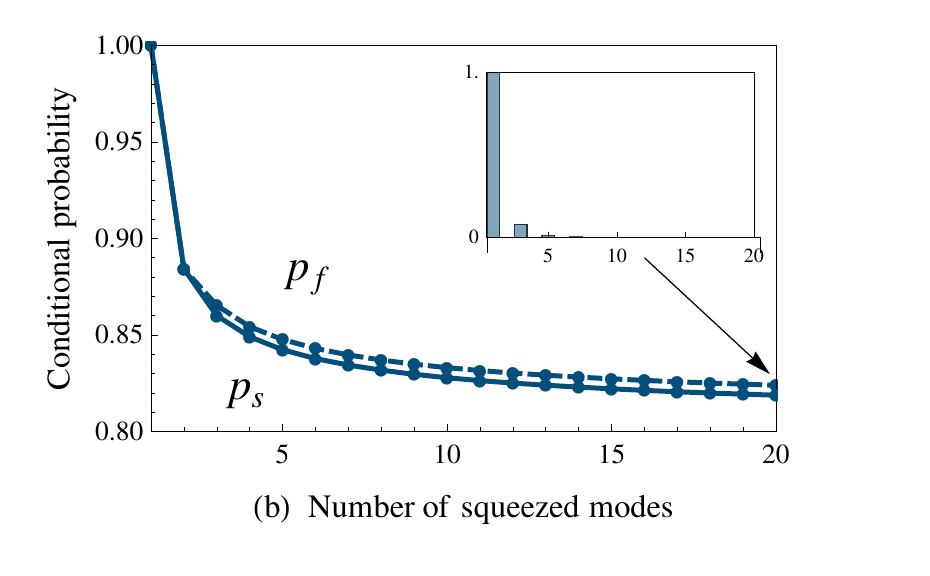}
	\caption{Conditional subtraction probability from the first Gaussian squeezed mode (solid line) after single-photon subtraction performed via weak parametric up-conversion with parameters depicted at Fig.\ref{fig:QPG_modes} and heralding probability of a single-photon state for weakly squeezed light (dashed line). Inset depicts the single-photon mode in the basis of squeezed modes. }
	\label{fig:purityQPG_single}
	\end{figure}

The conditional subtraction probability calculated from the first squeezed mode is higher than one-half level (see Fig.~\ref{fig:purityQPG_single}) and the Wigner function of the reduced state will have negative values.
It is due to the fact that the mode almost perfectly coincides with the first subtraction mode (defined by the Gaussian gate pulses) having the highest subtraction probability.
Also, the calculation performed in the weak-squeezing approximation (see dashed line in Fig.~\ref{fig:purityQPG_single} and the inset therein) shows that both the probability and the mode of the heralded single-photon state almost coincides with the first squeezed mode.

The aforementioned results show that both subtraction methods lead to the same purity of the multimode state under the experimentally relevant conditions.
However in the second case the probability to subtract a photon from the first Hermite-Gaussian mode is greater than $1/2$, while lesser in the first case.
The reason is the following.
In the first case photon is subtracted from a narrow band around central frequency that overlaps with $N/2$ even squeezed modes.
In the second case, there is almost unique overlap between the first squeezed mode and the subtraction mode.
The single-photon subtraction is then almost single-mode.
It is worth noticing that \citep{Averchenko2014} shows that the Schmidt number can be further decreased with a wider spectrum of the gate pulse thus leading to a purer single-photon subtraction.
Still, with a gate pulse matched to $u_0(\w)$, the achievable value of the subtraction probability is enough to obtain significant negativity in the state Wigner function.

Interestingly, the total subtraction probability $P$ allows one to assess the rate of subtraction events for the two methods. For example, for $N=10$ equally squeezed modes , a filter bandwidth of half the width of $u_0(\w)$ and a $1\%$ reflective beamsplitter, one could expect $P/n \approx 1.2 \times 10^{-2}$ per pulse for the linear scheme.
On the other hand, for the same number of equally squeezed modes, a gate pulse energy of $1nJ$ and a beam diameter of $1mm^2$, one could expect $P/n \approx 3.9 \times 10^{-4}$ per pulse in the non-linear apparatus.
Single-photon subtraction is thus about $30$ times more efficient with a simple beamsplitter than through non-linear interaction but the performance of the latter in terms of state purity and selectivity has been shown to be much better.

\subsection{Degaussification of a squeezed frequency comb}

We consider here the single-photon subtraction from a multimode squeezed frequency comb \cite{Roslund2013}.
Its highly multimode intrinsic structure and unique spectral and temporal modal properties are promising for measurement based scalable quantum computation \cite{MedeirosdeAraujo2014}.
As single squeezed pulse, such a comb is usually generated by a pump pulse undergoing parametric down-conversion in a non-linear medium, generally an optical crystal.
The whole optical non-linear process takes place in a optical cavity whose free spectral range governs the comb spectral structure.
The optical cavity allows resonant build up of the field inside it and leads to higher squeezing per unit pump power ($-4.2dB$ \cite{Roslund2013}) as compared with a single-pass configuration ($-2.0dB$ \cite{Slusher1987}).

Within the cavity, parametric down-conversion populates the optical spectrum with correlated signal and idler photons \cite{MedeirosdeAraujo2014}.
Because of energy conservation, those photons can only exist around different resonances of the optical cavity with respect to the cavity bandwidth.
A single squeezed mode in the frequency domain is a superposition of sidebands equally separated from their optical cavity resonances \cite{Averchenko2010,Jiang2012}.
This mode is independent from other sets of sidebands frequencies and hosts a squeezed quantum state.
The whole system is thus highly multimode. For a pump with Gaussian pulses, a single squeezed mode in the time domain is an infinite train of correlated Hermite-Gaussian pulses with a fixed phase relations between pulses.
In the time domain, the correlations between the same Hermite-Gaussian pulses are due to the optical cavity lifetime so that interpulse correlations are exponentially decaying.
The number of effectively correlated pulses is thus roughly given by the cavity finesse.

It is worth noticing that, under experimentally relevant conditions, the single-photon subtraction happens in a single pulse of a train.
The results obtained previously in Sec. \ref{pulses} for the purity of a conditioned multimode state and the subtraction probability hold for both subtraction protocols.
The first change is that $n_k$ becomes a number of photons per single pulse of a given mode, not per mode (see Appendix \ref{sec:appendix_comb}).
Moreover, the state of a squeezed mode after single-photon subtraction will be different from Sec. \ref{pulses}.
Indeed, as the single photon is subtracted from an infinite train, its impact on the embedded state will be negligible
Thus the Wigner function of the embedded state is not likely to be significantly affected and will not possesses any negativity.
In order to measure a non-Gaussian Wigner function, one has to experimentally isolate the temporal mode where the single-photon subtraction happens, just as in the continuous wave regime \cite{Morin2013}.

\begin{figure}[t]
	\center{\includegraphics[width=0.95\linewidth]{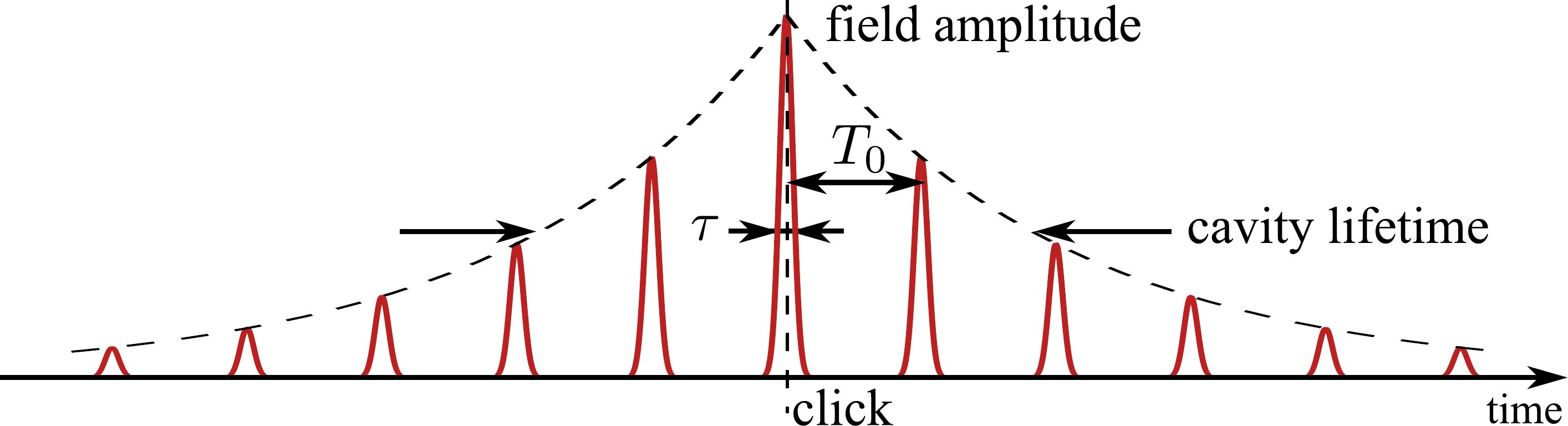}}
	\caption{Temporal mode of a single photon state conditioned on the single-photon subtraction from a weakly squeezed frequency comb. We assumed that the subtraction is performed from the first Hermite-Gaussian mode of a single pulse.}
	\label{heralded photon mode}
	\end{figure}

Finally, in the weak squeezing approximation, $N$ independent coherent superpositions of vacuum and a two-photon states are hosted by different single squeezed mode (just like the squeezed light generated by a continuously pumped optical parametric oscillator \cite{Nielsen2007}).
The detection of a single photon heralds a single-photon state. The mode where this single photon appears with the highest probability (being also the mode with the most negative Wigner function) can be estimated using the previous results and the treatment of the single-photon subtraction from continuous squeezed light \cite{Molmer2006, Nielsen2007, Morin2013}.
Then Figs.\ref{fig:BS_results_fock} and \ref{fig:purityQPG_single} show a mode where the photon is contained within a single pulse with the highest probability for the two different subtraction methods. The single-photon state is nevertheless still delocalized within a well defined temporal mode with an exponentially decaying profile.
This mode is centered in the time domain around the single-photon detection instant as illustrated on Fig.~\ref{heralded photon mode}.
The precise determination of this temporal profile is critical if ones wants to perform quantum state tomography through homodyne detection \cite{Morin2013}.

\section{Conclusion}\L{sec:concl}

We have introduced a general framework that allows to fully describe the subtraction of a single photon from a multimode quantum state through the computation of the process eigenmodes and eigenvalues.
The strength of this formalism is its adaptability to the modal description of the multimode state considered.
It can as well perfectly describe single-photon addition on similar resource.
We have shown that, given the modal nature of the state, of the splitting mechanism and of the single-photon detection, this framework predicts the properties of the single-photon subtracted multimode state.
One can then assess the selectivity of the single-photon subtraction or the entanglement introduced between modes that were at first independent.
As an example, we have applied this formalism for multimode squeezed vacuum in the weak squeezing approximation to determine the mode hosting a state that had the higher fidelity with a single-photon state.
In a more ambitious perspective, we have been able to fully determine the properties of a subtracted spectrally squeezed multimode state for both subtraction methods.
We have shown that a weak beamsplitter and a spectral filter result in a non-pure subtraction, thus readily degrading the multimode state purity to a point where no negativity can be observed in the Wigner function of the state of any mode.
On the other hand, subtracting a single photon via up-conversion engineered by a gate pulse allows one to keep a high overall state purity and even higher purity and Wigner function negativity for the target mode.
While a weak beamsplitter and a spectral filter constitute a simple and easily implemented solution compared to parametric up-conversion, its lack of selectivity makes this solution unreliable to fully engineer multimode quantum states.
On the contrary, controlling single-photon subtraction with a shaped gate beam offers wide possibilities of tuning the degaussification of multimode quantum states.

\acknowledgments

We thank Markus Grassl for useful discussions. V.A. acknowledges support of Ville de Paris and personal grant of the Max Planck Society.

\bibliography{biblio}

\begin{thebibliography}{48}%
\makeatletter
\providecommand \@ifxundefined [1]{%
 \@ifx{#1\undefined}
}%
\providecommand \@ifnum [1]{%
 \ifnum #1\expandafter \@firstoftwo
 \else \expandafter \@secondoftwo
 \fi
}%
\providecommand \@ifx [1]{%
 \ifx #1\expandafter \@firstoftwo
 \else \expandafter \@secondoftwo
 \fi
}%
\providecommand \natexlab [1]{#1}%
\providecommand \enquote  [1]{``#1''}%
\providecommand \bibnamefont  [1]{#1}%
\providecommand \bibfnamefont [1]{#1}%
\providecommand \citenamefont [1]{#1}%
\providecommand \href@noop [0]{\@secondoftwo}%
\providecommand \href [0]{\begingroup \@sanitize@url \@href}%
\providecommand \@href[1]{\@@startlink{#1}\@@href}%
\providecommand \@@href[1]{\endgroup#1\@@endlink}%
\providecommand \@sanitize@url [0]{\catcode `\\12\catcode `\$12\catcode
  `\&12\catcode `\#12\catcode `\^12\catcode `\_12\catcode `\%12\relax}%
\providecommand \@@startlink[1]{}%
\providecommand \@@endlink[0]{}%
\providecommand \url  [0]{\begingroup\@sanitize@url \@url }%
\providecommand \@url [1]{\endgroup\@href {#1}{\urlprefix }}%
\providecommand \urlprefix  [0]{URL }%
\providecommand \Eprint [0]{\href }%
\providecommand \doibase [0]{http://dx.doi.org/}%
\providecommand \selectlanguage [0]{\@gobble}%
\providecommand \bibinfo  [0]{\@secondoftwo}%
\providecommand \bibfield  [0]{\@secondoftwo}%
\providecommand \translation [1]{[#1]}%
\providecommand \BibitemOpen [0]{}%
\providecommand \bibitemStop [0]{}%
\providecommand \bibitemNoStop [0]{.\EOS\space}%
\providecommand \EOS [0]{\spacefactor3000\relax}%
\providecommand \BibitemShut  [1]{\csname bibitem#1\endcsname}%
\let\auto@bib@innerbib\@empty
\bibitem [{\citenamefont {O'Brien}\ \emph {et~al.}(2003)\citenamefont
  {O'Brien}, \citenamefont {Pryde}, \citenamefont {White}, \citenamefont
  {Ralph},\ and\ \citenamefont {Branning}}]{OBrien2003}%
  \BibitemOpen
  \bibfield  {author} {\bibinfo {author} {\bibfnamefont {J.~L.}\ \bibnamefont
  {O'Brien}}, \bibinfo {author} {\bibfnamefont {G.~J.}\ \bibnamefont {Pryde}},
  \bibinfo {author} {\bibfnamefont {a.~G.}\ \bibnamefont {White}}, \bibinfo
  {author} {\bibfnamefont {T.~C.}\ \bibnamefont {Ralph}}, \ and\ \bibinfo
  {author} {\bibfnamefont {D.}~\bibnamefont {Branning}},\ }\href {\doibase
  10.1038/natur=} {\bibfield  {journal} {\bibinfo  {journal} {Nature}\ }\textbf
  {\bibinfo {volume} {426}},\ \bibinfo {pages} {264} (\bibinfo {year}
  {2003})}\BibitemShut {NoStop}%
\bibitem [{\citenamefont {Roslund}\ \emph {et~al.}(2014)\citenamefont
  {Roslund}, \citenamefont {de~Ara\'{u}jo}, \citenamefont {Jiang},
  \citenamefont {Fabre},\ and\ \citenamefont {Treps}}]{Roslund2013}%
  \BibitemOpen
  \bibfield  {author} {\bibinfo {author} {\bibfnamefont {J.}~\bibnamefont
  {Roslund}}, \bibinfo {author} {\bibfnamefont {R.~M.}\ \bibnamefont
  {de~Ara\'{u}jo}}, \bibinfo {author} {\bibfnamefont {S.}~\bibnamefont
  {Jiang}}, \bibinfo {author} {\bibfnamefont {C.}~\bibnamefont {Fabre}}, \ and\
  \bibinfo {author} {\bibfnamefont {N.}~\bibnamefont {Treps}},\ }\href
  {\doibase 10.1038/nphoton.2013.340} {\bibfield  {journal} {\bibinfo
  {journal} {Nature Photonics}\ }\textbf {\bibinfo {volume} {8}},\ \bibinfo
  {pages} {109} (\bibinfo {year} {2014})}\BibitemShut {NoStop}%
\bibitem [{\citenamefont {Nielsen}\ and\ \citenamefont
  {M{\o}lmer}(2007)}]{Nielsen2007}%
  \BibitemOpen
  \bibfield  {author} {\bibinfo {author} {\bibfnamefont {A.}~\bibnamefont
  {Nielsen}}\ and\ \bibinfo {author} {\bibfnamefont {K.}~\bibnamefont
  {M{\o}lmer}},\ }\href {\doibase 10.1103/PhysRevA.75.023806} {\bibfield
  {journal} {\bibinfo  {journal} {Physical Review A}\ }\textbf {\bibinfo
  {volume} {75}},\ \bibinfo {pages} {023806} (\bibinfo {year}
  {2007})}\BibitemShut {NoStop}%
\bibitem [{\citenamefont {Yukawa}\ \emph
  {et~al.}(2013{\natexlab{a}})\citenamefont {Yukawa}, \citenamefont {Miyata},
  \citenamefont {Mizuta}, \citenamefont {Yonezawa}, \citenamefont {Marek},
  \citenamefont {Filip},\ and\ \citenamefont {Furusawa}}]{Yukawa2013a}%
  \BibitemOpen
  \bibfield  {author} {\bibinfo {author} {\bibfnamefont {M.}~\bibnamefont
  {Yukawa}}, \bibinfo {author} {\bibfnamefont {K.}~\bibnamefont {Miyata}},
  \bibinfo {author} {\bibfnamefont {T.}~\bibnamefont {Mizuta}}, \bibinfo
  {author} {\bibfnamefont {H.}~\bibnamefont {Yonezawa}}, \bibinfo {author}
  {\bibfnamefont {P.}~\bibnamefont {Marek}}, \bibinfo {author} {\bibfnamefont
  {R.}~\bibnamefont {Filip}}, \ and\ \bibinfo {author} {\bibfnamefont
  {A.}~\bibnamefont {Furusawa}},\ }\href {\doibase 10.1364/OE.21.005529}
  {\bibfield  {journal} {\bibinfo  {journal} {Optics Express}\ }\textbf
  {\bibinfo {volume} {21}},\ \bibinfo {pages} {5} (\bibinfo {year}
  {2013}{\natexlab{a}})}\BibitemShut {NoStop}%
\bibitem [{\citenamefont {Ourjoumtsev}\ \emph {et~al.}(2006)\citenamefont
  {Ourjoumtsev}, \citenamefont {Tualle-Brouri}, \citenamefont {Laurat},\ and\
  \citenamefont {Grangier}}]{Ourjoumtsev2006}%
  \BibitemOpen
  \bibfield  {author} {\bibinfo {author} {\bibfnamefont {A.}~\bibnamefont
  {Ourjoumtsev}}, \bibinfo {author} {\bibfnamefont {R.}~\bibnamefont
  {Tualle-Brouri}}, \bibinfo {author} {\bibfnamefont {J.}~\bibnamefont
  {Laurat}}, \ and\ \bibinfo {author} {\bibfnamefont {P.}~\bibnamefont
  {Grangier}},\ }\href {http://www.sciencemag.org/content/312/5770/83.short}
  {\bibfield  {journal} {\bibinfo  {journal} {Science}\ }\textbf {\bibinfo
  {volume} {312}},\ \bibinfo {pages} {83} (\bibinfo {year} {2006})}\BibitemShut
  {NoStop}%
\bibitem [{\citenamefont {Morin}\ \emph {et~al.}(2014)\citenamefont {Morin},
  \citenamefont {Huang}, \citenamefont {Liu}, \citenamefont {{Le Jeannic}},
  \citenamefont {Fabre},\ and\ \citenamefont {Laurat}}]{Morin2014}%
  \BibitemOpen
  \bibfield  {author} {\bibinfo {author} {\bibfnamefont {O.}~\bibnamefont
  {Morin}}, \bibinfo {author} {\bibfnamefont {K.}~\bibnamefont {Huang}},
  \bibinfo {author} {\bibfnamefont {J.}~\bibnamefont {Liu}}, \bibinfo {author}
  {\bibfnamefont {H.}~\bibnamefont {{Le Jeannic}}}, \bibinfo {author}
  {\bibfnamefont {C.}~\bibnamefont {Fabre}}, \ and\ \bibinfo {author}
  {\bibfnamefont {J.}~\bibnamefont {Laurat}},\ }\href {\doibase
  10.1038/nphoton.2014.137} {\bibfield  {journal} {\bibinfo  {journal} {Nature
  Photonics}\ }\textbf {\bibinfo {volume} {8}},\ \bibinfo {pages} {570}
  (\bibinfo {year} {2014})}\BibitemShut {NoStop}%
\bibitem [{\citenamefont {Ourjoumtsev}\ \emph {et~al.}(2009)\citenamefont
  {Ourjoumtsev}, \citenamefont {Ferreyrol}, \citenamefont {Tualle-Brouri},\
  and\ \citenamefont {Grangier}}]{Ourjoumtsev2009}%
  \BibitemOpen
  \bibfield  {author} {\bibinfo {author} {\bibfnamefont {A.}~\bibnamefont
  {Ourjoumtsev}}, \bibinfo {author} {\bibfnamefont {F.}~\bibnamefont
  {Ferreyrol}}, \bibinfo {author} {\bibfnamefont {R.}~\bibnamefont
  {Tualle-Brouri}}, \ and\ \bibinfo {author} {\bibfnamefont {P.}~\bibnamefont
  {Grangier}},\ }\href {\doibase 10.1038/nphys1199} {\bibfield  {journal}
  {\bibinfo  {journal} {Nature Physics}\ }\textbf {\bibinfo {volume} {5}},\
  \bibinfo {pages} {189} (\bibinfo {year} {2009})}\BibitemShut {NoStop}%
\bibitem [{\citenamefont {Takahashi}\ \emph {et~al.}(2010)\citenamefont
  {Takahashi}, \citenamefont {Neergaard-Nielsen}, \citenamefont {Takeuchi},
  \citenamefont {Takeoka}, \citenamefont {Hayasaka}, \citenamefont {Furusawa},\
  and\ \citenamefont {Sasaki}}]{Takahashi2010}%
  \BibitemOpen
  \bibfield  {author} {\bibinfo {author} {\bibfnamefont {H.}~\bibnamefont
  {Takahashi}}, \bibinfo {author} {\bibfnamefont {J.~S.}\ \bibnamefont
  {Neergaard-Nielsen}}, \bibinfo {author} {\bibfnamefont {M.}~\bibnamefont
  {Takeuchi}}, \bibinfo {author} {\bibfnamefont {M.}~\bibnamefont {Takeoka}},
  \bibinfo {author} {\bibfnamefont {K.}~\bibnamefont {Hayasaka}}, \bibinfo
  {author} {\bibfnamefont {A.}~\bibnamefont {Furusawa}}, \ and\ \bibinfo
  {author} {\bibfnamefont {M.}~\bibnamefont {Sasaki}},\ }\href {\doibase
  10.1038/nphoton.2010.1} {\bibfield  {journal} {\bibinfo  {journal} {Nature
  Photonics}\ }\textbf {\bibinfo {volume} {4}},\ \bibinfo {pages} {8} (\bibinfo
  {year} {2010})}\BibitemShut {NoStop}%
\bibitem [{\citenamefont {Parigi}\ \emph {et~al.}(2007)\citenamefont {Parigi},
  \citenamefont {Zavatta}, \citenamefont {Kim},\ and\ \citenamefont
  {Bellini}}]{Parigi2007}%
  \BibitemOpen
  \bibfield  {author} {\bibinfo {author} {\bibfnamefont {V.}~\bibnamefont
  {Parigi}}, \bibinfo {author} {\bibfnamefont {A.}~\bibnamefont {Zavatta}},
  \bibinfo {author} {\bibfnamefont {M.}~\bibnamefont {Kim}}, \ and\ \bibinfo
  {author} {\bibfnamefont {M.}~\bibnamefont {Bellini}},\ }\href {\doibase
  10.1126/science.1146204} {\bibfield  {journal} {\bibinfo  {journal}
  {Science}\ }\textbf {\bibinfo {volume} {317}},\ \bibinfo {pages} {1890}
  (\bibinfo {year} {2007})}\BibitemShut {NoStop}%
\bibitem [{\citenamefont {Marek}\ \emph {et~al.}(2011)\citenamefont {Marek},
  \citenamefont {Filip},\ and\ \citenamefont {Furusawa}}]{Marek2011}%
  \BibitemOpen
  \bibfield  {author} {\bibinfo {author} {\bibfnamefont {P.}~\bibnamefont
  {Marek}}, \bibinfo {author} {\bibfnamefont {R.}~\bibnamefont {Filip}}, \ and\
  \bibinfo {author} {\bibfnamefont {A.}~\bibnamefont {Furusawa}},\ }\href
  {\doibase 10.1103/PhysRevA.84.053802} {\bibfield  {journal} {\bibinfo
  {journal} {Physical Review A}\ }\textbf {\bibinfo {volume} {84}},\ \bibinfo
  {pages} {1} (\bibinfo {year} {2011})}\BibitemShut {NoStop}%
\bibitem [{\citenamefont {Yukawa}\ \emph
  {et~al.}(2013{\natexlab{b}})\citenamefont {Yukawa}, \citenamefont {Miyata},
  \citenamefont {Yonezawa}, \citenamefont {Marek}, \citenamefont {Filip},\ and\
  \citenamefont {Furusawa}}]{Yukawa2013b}%
  \BibitemOpen
  \bibfield  {author} {\bibinfo {author} {\bibfnamefont {M.}~\bibnamefont
  {Yukawa}}, \bibinfo {author} {\bibfnamefont {K.}~\bibnamefont {Miyata}},
  \bibinfo {author} {\bibfnamefont {H.}~\bibnamefont {Yonezawa}}, \bibinfo
  {author} {\bibfnamefont {P.}~\bibnamefont {Marek}}, \bibinfo {author}
  {\bibfnamefont {R.}~\bibnamefont {Filip}}, \ and\ \bibinfo {author}
  {\bibfnamefont {A.}~\bibnamefont {Furusawa}},\ }\href {\doibase
  10.1103/PhysRevA.88.053816} {\bibfield  {journal} {\bibinfo  {journal}
  {Physical Review A}\ }\textbf {\bibinfo {volume} {88}},\ \bibinfo {pages} {1}
  (\bibinfo {year} {2013}{\natexlab{b}})}\BibitemShut {NoStop}%
\bibitem [{\citenamefont {Marshall}\ \emph {et~al.}(2014)\citenamefont
  {Marshall}, \citenamefont {Pooser}, \citenamefont {Siopsis},\ and\
  \citenamefont {Weedbrook}}]{Marshall2014}%
  \BibitemOpen
  \bibfield  {author} {\bibinfo {author} {\bibfnamefont {K.}~\bibnamefont
  {Marshall}}, \bibinfo {author} {\bibfnamefont {R.}~\bibnamefont {Pooser}},
  \bibinfo {author} {\bibfnamefont {G.}~\bibnamefont {Siopsis}}, \ and\
  \bibinfo {author} {\bibfnamefont {C.}~\bibnamefont {Weedbrook}},\ }\href
  {http://arxiv.org/abs/1412.0336v1} {\ \textbf {\bibinfo {volume} {2}},\
  \bibinfo {pages} {1} (\bibinfo {year} {2014})},\ \Eprint
  {http://arxiv.org/abs/1412.0336} {arXiv:1412.0336} \BibitemShut {NoStop}%
\bibitem [{\citenamefont {Olson}\ \emph {et~al.}(2015)\citenamefont {Olson},
  \citenamefont {Seshadreesan}, \citenamefont {Motes}, \citenamefont {Rohde},\
  and\ \citenamefont {Dowling}}]{Olson2015}%
  \BibitemOpen
  \bibfield  {author} {\bibinfo {author} {\bibfnamefont {J.~P.}\ \bibnamefont
  {Olson}}, \bibinfo {author} {\bibfnamefont {K.~P.}\ \bibnamefont
  {Seshadreesan}}, \bibinfo {author} {\bibfnamefont {K.~R.}\ \bibnamefont
  {Motes}}, \bibinfo {author} {\bibfnamefont {P.~P.}\ \bibnamefont {Rohde}}, \
  and\ \bibinfo {author} {\bibfnamefont {J.~P.}\ \bibnamefont {Dowling}},\
  }\href {\doibase 10.1103/PhysRevA.91.022317} {\bibfield  {journal} {\bibinfo
  {journal} {Physical Review A}\ }\textbf {\bibinfo {volume} {91}},\ \bibinfo
  {pages} {022317} (\bibinfo {year} {2015})}\BibitemShut {NoStop}%
\bibitem [{\citenamefont {Tualle-Brouri}\ \emph {et~al.}(2009)\citenamefont
  {Tualle-Brouri}, \citenamefont {Ourjoumtsev}, \citenamefont {Dantan},
  \citenamefont {Grangier}, \citenamefont {Wubs},\ and\ \citenamefont
  {Sorensen}}]{Tualle-Brouri2009}%
  \BibitemOpen
  \bibfield  {author} {\bibinfo {author} {\bibfnamefont {R.}~\bibnamefont
  {Tualle-Brouri}}, \bibinfo {author} {\bibfnamefont {A.}~\bibnamefont
  {Ourjoumtsev}}, \bibinfo {author} {\bibfnamefont {A.}~\bibnamefont {Dantan}},
  \bibinfo {author} {\bibfnamefont {P.}~\bibnamefont {Grangier}}, \bibinfo
  {author} {\bibfnamefont {M.}~\bibnamefont {Wubs}}, \ and\ \bibinfo {author}
  {\bibfnamefont {A.}~\bibnamefont {Sorensen}},\ }\href {\doibase
  10.1103/PhysRevA.80.013806} {\bibfield  {journal} {\bibinfo  {journal}
  {Physical Review A}\ }\textbf {\bibinfo {volume} {80}},\ \bibinfo {pages}
  {013806} (\bibinfo {year} {2009})}\BibitemShut {NoStop}%
\bibitem [{\citenamefont {Takahashi}\ \emph {et~al.}(2008)\citenamefont
  {Takahashi}, \citenamefont {Wakui}, \citenamefont {Suzuki}, \citenamefont
  {Takeoka}, \citenamefont {Hayasaka}, \citenamefont {Furusawa},\ and\
  \citenamefont {Sasaki}}]{Takahashi2008}%
  \BibitemOpen
  \bibfield  {author} {\bibinfo {author} {\bibfnamefont {H.}~\bibnamefont
  {Takahashi}}, \bibinfo {author} {\bibfnamefont {K.}~\bibnamefont {Wakui}},
  \bibinfo {author} {\bibfnamefont {S.}~\bibnamefont {Suzuki}}, \bibinfo
  {author} {\bibfnamefont {M.}~\bibnamefont {Takeoka}}, \bibinfo {author}
  {\bibfnamefont {K.}~\bibnamefont {Hayasaka}}, \bibinfo {author}
  {\bibfnamefont {A.}~\bibnamefont {Furusawa}}, \ and\ \bibinfo {author}
  {\bibfnamefont {M.}~\bibnamefont {Sasaki}},\ }\href {\doibase
  10.1103/PhysRevLett.101.233605} {\bibfield  {journal} {\bibinfo  {journal}
  {Physical Review Letters}\ }\textbf {\bibinfo {volume} {101}},\ \bibinfo
  {pages} {16} (\bibinfo {year} {2008})}\BibitemShut {NoStop}%
\bibitem [{\citenamefont {Su}\ \emph {et~al.}(2007)\citenamefont {Su},
  \citenamefont {Tan}, \citenamefont {Jia}, \citenamefont {Zhang},
  \citenamefont {Xie},\ and\ \citenamefont {Peng}}]{Su2007}%
  \BibitemOpen
  \bibfield  {author} {\bibinfo {author} {\bibfnamefont {X.}~\bibnamefont
  {Su}}, \bibinfo {author} {\bibfnamefont {A.}~\bibnamefont {Tan}}, \bibinfo
  {author} {\bibfnamefont {X.}~\bibnamefont {Jia}}, \bibinfo {author}
  {\bibfnamefont {J.}~\bibnamefont {Zhang}}, \bibinfo {author} {\bibfnamefont
  {C.}~\bibnamefont {Xie}}, \ and\ \bibinfo {author} {\bibfnamefont
  {K.}~\bibnamefont {Peng}},\ }\href {\doibase 10.1103/PhysRevLett.98.070502}
  {\bibfield  {journal} {\bibinfo  {journal} {Physical Review Letters}\
  }\textbf {\bibinfo {volume} {98}},\ \bibinfo {pages} {2} (\bibinfo {year}
  {2007})}\BibitemShut {NoStop}%
\bibitem [{\citenamefont {Yukawa}\ \emph {et~al.}(2008)\citenamefont {Yukawa},
  \citenamefont {Ukai}, \citenamefont {van Loock},\ and\ \citenamefont
  {Furusawa}}]{Yukawa2008}%
  \BibitemOpen
  \bibfield  {author} {\bibinfo {author} {\bibfnamefont {M.}~\bibnamefont
  {Yukawa}}, \bibinfo {author} {\bibfnamefont {R.}~\bibnamefont {Ukai}},
  \bibinfo {author} {\bibfnamefont {P.}~\bibnamefont {van Loock}}, \ and\
  \bibinfo {author} {\bibfnamefont {A.}~\bibnamefont {Furusawa}},\ }\href
  {\doibase 10.1103/PhysRevA.78.012301} {\bibfield  {journal} {\bibinfo
  {journal} {Physical Review A}\ }\textbf {\bibinfo {volume} {78}},\ \bibinfo
  {pages} {012301} (\bibinfo {year} {2008})}\BibitemShut {NoStop}%
\bibitem [{\citenamefont {{Medeiros de Ara\'{u}jo}}\ \emph
  {et~al.}(2014)\citenamefont {{Medeiros de Ara\'{u}jo}}, \citenamefont
  {Roslund}, \citenamefont {Cai}, \citenamefont {Ferrini}, \citenamefont
  {Fabre},\ and\ \citenamefont {Treps}}]{MedeirosdeAraujo2014}%
  \BibitemOpen
  \bibfield  {author} {\bibinfo {author} {\bibfnamefont {R.}~\bibnamefont
  {{Medeiros de Ara\'{u}jo}}}, \bibinfo {author} {\bibfnamefont
  {J.}~\bibnamefont {Roslund}}, \bibinfo {author} {\bibfnamefont
  {Y.}~\bibnamefont {Cai}}, \bibinfo {author} {\bibfnamefont {G.}~\bibnamefont
  {Ferrini}}, \bibinfo {author} {\bibfnamefont {C.}~\bibnamefont {Fabre}}, \
  and\ \bibinfo {author} {\bibfnamefont {N.}~\bibnamefont {Treps}},\ }\href
  {\doibase 10.1103/PhysRevA.89.053828} {\bibfield  {journal} {\bibinfo
  {journal} {Physical Review A}\ }\textbf {\bibinfo {volume} {89}},\ \bibinfo
  {pages} {053828} (\bibinfo {year} {2014})}\BibitemShut {NoStop}%
\bibitem [{\citenamefont {Chen}\ \emph {et~al.}(2014)\citenamefont {Chen},
  \citenamefont {Menicucci},\ and\ \citenamefont {Pfister}}]{Chen2014}%
  \BibitemOpen
  \bibfield  {author} {\bibinfo {author} {\bibfnamefont {M.}~\bibnamefont
  {Chen}}, \bibinfo {author} {\bibfnamefont {N.~C.}\ \bibnamefont {Menicucci}},
  \ and\ \bibinfo {author} {\bibfnamefont {O.}~\bibnamefont {Pfister}},\ }\href
  {\doibase 10.1103/PhysRevLett.112.120505} {\bibfield  {journal} {\bibinfo
  {journal} {Physical Review Letters}\ }\textbf {\bibinfo {volume} {112}},\
  \bibinfo {pages} {120505} (\bibinfo {year} {2014})}\BibitemShut {NoStop}%
\bibitem [{\citenamefont {Ekert}\ and\ \citenamefont
  {Knight}(1995)}]{Ekert1995}%
  \BibitemOpen
  \bibfield  {author} {\bibinfo {author} {\bibfnamefont {A.}~\bibnamefont
  {Ekert}}\ and\ \bibinfo {author} {\bibfnamefont {P.}~\bibnamefont {Knight}},\
  }\href {\doibase 10.1119/1.17904} {\bibfield  {journal} {\bibinfo  {journal}
  {American Journal of Physics}\ }\textbf {\bibinfo {volume} {63}},\ \bibinfo
  {pages} {415} (\bibinfo {year} {1995})}\BibitemShut {NoStop}%
\bibitem [{\citenamefont {Dakna}\ \emph {et~al.}(1997)\citenamefont {Dakna},
  \citenamefont {Anhut}, \citenamefont {Opatrn\'{y}}, \citenamefont
  {Kn\"{o}ll},\ and\ \citenamefont {Welsch}}]{Dakna1996}%
  \BibitemOpen
  \bibfield  {author} {\bibinfo {author} {\bibfnamefont {M.}~\bibnamefont
  {Dakna}}, \bibinfo {author} {\bibfnamefont {T.}~\bibnamefont {Anhut}},
  \bibinfo {author} {\bibfnamefont {T.}~\bibnamefont {Opatrn\'{y}}}, \bibinfo
  {author} {\bibfnamefont {L.}~\bibnamefont {Kn\"{o}ll}}, \ and\ \bibinfo
  {author} {\bibfnamefont {D.~G.}\ \bibnamefont {Welsch}},\ }\href
  {http://pra.aps.org/abstract/PRA/v55/i4/p3184\_1} {\bibfield  {journal}
  {\bibinfo  {journal} {Physical Review A}\ }\textbf {\bibinfo {volume} {55}},\
  \bibinfo {pages} {3184} (\bibinfo {year} {1997})}\BibitemShut {NoStop}%
\bibitem [{\citenamefont {Eckstein}\ \emph {et~al.}(2011)\citenamefont
  {Eckstein}, \citenamefont {Brecht},\ and\ \citenamefont
  {Silberhorn}}]{Eckstein2011}%
  \BibitemOpen
  \bibfield  {author} {\bibinfo {author} {\bibfnamefont {A.}~\bibnamefont
  {Eckstein}}, \bibinfo {author} {\bibfnamefont {B.}~\bibnamefont {Brecht}}, \
  and\ \bibinfo {author} {\bibfnamefont {C.}~\bibnamefont {Silberhorn}},\
  }\href {http://www.ncbi.nlm.nih.gov/pubmed/21934737} {\bibfield  {journal}
  {\bibinfo  {journal} {Optics express}\ }\textbf {\bibinfo {volume} {19}},\
  \bibinfo {pages} {13770} (\bibinfo {year} {2011})}\BibitemShut {NoStop}%
\bibitem [{\citenamefont {Averchenko}\ \emph {et~al.}(2014)\citenamefont
  {Averchenko}, \citenamefont {Thiel},\ and\ \citenamefont
  {Treps}}]{Averchenko2014}%
  \BibitemOpen
  \bibfield  {author} {\bibinfo {author} {\bibfnamefont {V.}~\bibnamefont
  {Averchenko}}, \bibinfo {author} {\bibfnamefont {V.}~\bibnamefont {Thiel}}, \
  and\ \bibinfo {author} {\bibfnamefont {N.}~\bibnamefont {Treps}},\ }\href
  {\doibase 10.1103/PhysRevA.89.063808} {\bibfield  {journal} {\bibinfo
  {journal} {Physical Review A}\ }\textbf {\bibinfo {volume} {89}},\ \bibinfo
  {pages} {063808} (\bibinfo {year} {2014})}\BibitemShut {NoStop}%
\bibitem [{\citenamefont {Slusher}\ \emph {et~al.}(1987)\citenamefont
  {Slusher}, \citenamefont {Grangier}, \citenamefont {LaPorta}, \citenamefont
  {Yurke},\ and\ \citenamefont {Potasek}}]{Slusher1987}%
  \BibitemOpen
  \bibfield  {author} {\bibinfo {author} {\bibfnamefont {R.~E.}\ \bibnamefont
  {Slusher}}, \bibinfo {author} {\bibfnamefont {P.}~\bibnamefont {Grangier}},
  \bibinfo {author} {\bibfnamefont {A.}~\bibnamefont {LaPorta}}, \bibinfo
  {author} {\bibfnamefont {B.}~\bibnamefont {Yurke}}, \ and\ \bibinfo {author}
  {\bibfnamefont {M.~J.}\ \bibnamefont {Potasek}},\ }\href {\doibase
  10.1103/PhysRevLett.59.2566} {\bibfield  {journal} {\bibinfo  {journal}
  {Physical Review Letters}\ }\textbf {\bibinfo {volume} {59}},\ \bibinfo
  {pages} {2566} (\bibinfo {year} {1987})}\BibitemShut {NoStop}%
\bibitem [{\citenamefont {Wenger}\ \emph {et~al.}(2004)\citenamefont {Wenger},
  \citenamefont {Tualle-Brouri},\ and\ \citenamefont {Grangier}}]{Wenger2004}%
  \BibitemOpen
  \bibfield  {author} {\bibinfo {author} {\bibfnamefont {J.}~\bibnamefont
  {Wenger}}, \bibinfo {author} {\bibfnamefont {R.}~\bibnamefont
  {Tualle-Brouri}}, \ and\ \bibinfo {author} {\bibfnamefont {P.}~\bibnamefont
  {Grangier}},\ }\href {\doibase 10.1103/PhysRevLett.92.153601} {\bibfield
  {journal} {\bibinfo  {journal} {Physical Review Letters}\ }\textbf {\bibinfo
  {volume} {92}},\ \bibinfo {pages} {153601} (\bibinfo {year}
  {2004})}\BibitemShut {NoStop}%
\bibitem [{\citenamefont {Kolobov}(1999)}]{Kolobov1999}%
  \BibitemOpen
  \bibfield  {author} {\bibinfo {author} {\bibfnamefont {M.}~\bibnamefont
  {Kolobov}},\ }\href {\doibase 10.1103/RevModPhys.71.1539} {\bibfield
  {journal} {\bibinfo  {journal} {Reviews of Modern Physics}\ }\textbf
  {\bibinfo {volume} {71}},\ \bibinfo {pages} {1539} (\bibinfo {year}
  {1999})}\BibitemShut {NoStop}%
\bibitem [{\citenamefont {Treps}\ \emph {et~al.}(2005)\citenamefont {Treps},
  \citenamefont {Delaubert}, \citenamefont {Ma\^{\i}tre}, \citenamefont
  {Courty},\ and\ \citenamefont {Fabre}}]{Treps2005}%
  \BibitemOpen
  \bibfield  {author} {\bibinfo {author} {\bibfnamefont {N.}~\bibnamefont
  {Treps}}, \bibinfo {author} {\bibfnamefont {V.}~\bibnamefont {Delaubert}},
  \bibinfo {author} {\bibfnamefont {A.}~\bibnamefont {Ma\^{\i}tre}}, \bibinfo
  {author} {\bibfnamefont {J.}~\bibnamefont {Courty}}, \ and\ \bibinfo {author}
  {\bibfnamefont {C.}~\bibnamefont {Fabre}},\ }\href {\doibase
  10.1103/PhysRevA.71.013820} {\bibfield  {journal} {\bibinfo  {journal}
  {Physical Review A}\ }\textbf {\bibinfo {volume} {71}},\ \bibinfo {pages}
  {013820} (\bibinfo {year} {2005})}\BibitemShut {NoStop}%
\bibitem [{\citenamefont {Barnett}(2009)}]{Barnett2009}%
  \BibitemOpen
  \bibfield  {author} {\bibinfo {author} {\bibfnamefont {S.~M.}\ \bibnamefont
  {Barnett}},\ }\href {http://www.dfi.uchile.cl/~rsoto/programas/FI6670.pdf
  http://adsabs.harvard.edu/abs/2001qieg.conf.....A} {\emph {\bibinfo {title}
  {Quantum Information}}}\ (\bibinfo {year} {2009})\BibitemShut {NoStop}%
\bibitem [{\citenamefont {Ourjoumtsev}\ \emph {et~al.}(2007)\citenamefont
  {Ourjoumtsev}, \citenamefont {Dantan}, \citenamefont {Tualle-Brouri},\ and\
  \citenamefont {Grangier}}]{Ourjoumtsev2007}%
  \BibitemOpen
  \bibfield  {author} {\bibinfo {author} {\bibfnamefont {A.}~\bibnamefont
  {Ourjoumtsev}}, \bibinfo {author} {\bibfnamefont {A.}~\bibnamefont {Dantan}},
  \bibinfo {author} {\bibfnamefont {R.}~\bibnamefont {Tualle-Brouri}}, \ and\
  \bibinfo {author} {\bibfnamefont {P.}~\bibnamefont {Grangier}},\ }\href
  {\doibase 10.1103/PhysRevLett.98.030502} {\bibfield  {journal} {\bibinfo
  {journal} {Physical Review Letters}\ }\textbf {\bibinfo {volume} {98}},\
  \bibinfo {pages} {030502} (\bibinfo {year} {2007})}\BibitemShut {NoStop}%
\bibitem [{\citenamefont {Kurochkin}\ \emph {et~al.}(2014)\citenamefont
  {Kurochkin}, \citenamefont {Prasad},\ and\ \citenamefont
  {Lvovsky}}]{Kurochkin2014}%
  \BibitemOpen
  \bibfield  {author} {\bibinfo {author} {\bibfnamefont {Y.}~\bibnamefont
  {Kurochkin}}, \bibinfo {author} {\bibfnamefont {A.~S.}\ \bibnamefont
  {Prasad}}, \ and\ \bibinfo {author} {\bibfnamefont {a.~I.}\ \bibnamefont
  {Lvovsky}},\ }\href {\doibase 10.1103/PhysRevLett.112.070402} {\bibfield
  {journal} {\bibinfo  {journal} {Physical Review Letters}\ }\textbf {\bibinfo
  {volume} {112}},\ \bibinfo {pages} {070402} (\bibinfo {year}
  {2014})}\BibitemShut {NoStop}%
\bibitem [{\citenamefont {Yokoyama}\ \emph {et~al.}(2013)\citenamefont
  {Yokoyama}, \citenamefont {Ukai}, \citenamefont {Armstrong}, \citenamefont
  {Sornphiphatphong}, \citenamefont {Kaji}, \citenamefont {Suzuki},
  \citenamefont {Yoshikawa}, \citenamefont {Yonezawa}, \citenamefont
  {Menicucci},\ and\ \citenamefont {Furusawa}}]{Yokoyama2013}%
  \BibitemOpen
  \bibfield  {author} {\bibinfo {author} {\bibfnamefont {S.}~\bibnamefont
  {Yokoyama}}, \bibinfo {author} {\bibfnamefont {R.}~\bibnamefont {Ukai}},
  \bibinfo {author} {\bibfnamefont {S.~C.}\ \bibnamefont {Armstrong}}, \bibinfo
  {author} {\bibfnamefont {C.}~\bibnamefont {Sornphiphatphong}}, \bibinfo
  {author} {\bibfnamefont {T.}~\bibnamefont {Kaji}}, \bibinfo {author}
  {\bibfnamefont {S.}~\bibnamefont {Suzuki}}, \bibinfo {author} {\bibfnamefont
  {J.-i.}\ \bibnamefont {Yoshikawa}}, \bibinfo {author} {\bibfnamefont
  {H.}~\bibnamefont {Yonezawa}}, \bibinfo {author} {\bibfnamefont {N.~C.}\
  \bibnamefont {Menicucci}}, \ and\ \bibinfo {author} {\bibfnamefont
  {A.}~\bibnamefont {Furusawa}},\ }\href {\doibase 10.1038/nphoton.2013.287}
  {\bibfield  {journal} {\bibinfo  {journal} {Nature Photonics}\ }\textbf
  {\bibinfo {volume} {7}},\ \bibinfo {pages} {982} (\bibinfo {year}
  {2013})}\BibitemShut {NoStop}%
\bibitem [{\citenamefont {Jozsa}(1994)}]{Jozsa1994}%
  \BibitemOpen
  \bibfield  {author} {\bibinfo {author} {\bibfnamefont {R.}~\bibnamefont
  {Jozsa}},\ }\href {\doibase 10.1080/09500349414552171} {\bibfield  {journal}
  {\bibinfo  {journal} {Journal of Modern Optics}\ }\textbf {\bibinfo {volume}
  {41}},\ \bibinfo {pages} {2315} (\bibinfo {year} {1994})}\BibitemShut
  {NoStop}%
\bibitem [{\citenamefont {Biswas}\ and\ \citenamefont
  {Agarwal}(2007)}]{Biswas2007}%
  \BibitemOpen
  \bibfield  {author} {\bibinfo {author} {\bibfnamefont {A.}~\bibnamefont
  {Biswas}}\ and\ \bibinfo {author} {\bibfnamefont {G.}~\bibnamefont
  {Agarwal}},\ }\href {\doibase 10.1103/PhysRevA.75.032104} {\bibfield
  {journal} {\bibinfo  {journal} {Physical Review A}\ }\textbf {\bibinfo
  {volume} {75}},\ \bibinfo {pages} {032104} (\bibinfo {year}
  {2007})}\BibitemShut {NoStop}%
\bibitem [{\citenamefont {Neergaard-Nielsen}\ \emph {et~al.}(2007)\citenamefont
  {Neergaard-Nielsen}, \citenamefont {Nielsen}, \citenamefont {Takahashi},
  \citenamefont {Vistnes},\ and\ \citenamefont
  {Polzik}}]{Neergaard-Nielsen2007}%
  \BibitemOpen
  \bibfield  {author} {\bibinfo {author} {\bibfnamefont {J.~S.}\ \bibnamefont
  {Neergaard-Nielsen}}, \bibinfo {author} {\bibfnamefont {B.~M.}\ \bibnamefont
  {Nielsen}}, \bibinfo {author} {\bibfnamefont {H.}~\bibnamefont {Takahashi}},
  \bibinfo {author} {\bibfnamefont {a.~I.}\ \bibnamefont {Vistnes}}, \ and\
  \bibinfo {author} {\bibfnamefont {E.~S.}\ \bibnamefont {Polzik}},\ }\href
  {http://www.ncbi.nlm.nih.gov/pubmed/19547121} {\bibfield  {journal} {\bibinfo
   {journal} {Optics express}\ }\textbf {\bibinfo {volume} {15}},\ \bibinfo
  {pages} {7940} (\bibinfo {year} {2007})}\BibitemShut {NoStop}%
\bibitem [{\citenamefont {M{\o}lmer}(2006)}]{Molmer2006}%
  \BibitemOpen
  \bibfield  {author} {\bibinfo {author} {\bibfnamefont {K.}~\bibnamefont
  {M{\o}lmer}},\ }\href {\doibase 10.1103/PhysRevA.73.063804} {\bibfield
  {journal} {\bibinfo  {journal} {Physical Review A}\ }\textbf {\bibinfo
  {volume} {73}},\ \bibinfo {pages} {063804} (\bibinfo {year}
  {2006})}\BibitemShut {NoStop}%
\bibitem [{\citenamefont {Morin}\ \emph {et~al.}(2013)\citenamefont {Morin},
  \citenamefont {Fabre},\ and\ \citenamefont {Laurat}}]{Morin2013}%
  \BibitemOpen
  \bibfield  {author} {\bibinfo {author} {\bibfnamefont {O.}~\bibnamefont
  {Morin}}, \bibinfo {author} {\bibfnamefont {C.}~\bibnamefont {Fabre}}, \ and\
  \bibinfo {author} {\bibfnamefont {J.}~\bibnamefont {Laurat}},\ }\href
  {\doibase 10.1103/PhysRevLett.111.213602} {\bibfield  {journal} {\bibinfo
  {journal} {Physical Review Letters}\ }\textbf {\bibinfo {volume} {111}},\
  \bibinfo {pages} {213602} (\bibinfo {year} {2013})}\BibitemShut {NoStop}%
\bibitem [{\citenamefont {Law}\ \emph {et~al.}(2000)\citenamefont {Law},
  \citenamefont {Walmsley},\ and\ \citenamefont {Eberly}}]{Law2000}%
  \BibitemOpen
  \bibfield  {author} {\bibinfo {author} {\bibfnamefont {C.}~\bibnamefont
  {Law}}, \bibinfo {author} {\bibfnamefont {I.}~\bibnamefont {Walmsley}}, \
  and\ \bibinfo {author} {\bibfnamefont {J.}~\bibnamefont {Eberly}},\ }\href
  {http://www.ncbi.nlm.nih.gov/pubmed/10990929} {\bibfield  {journal} {\bibinfo
   {journal} {Physical Review Letters}\ }\textbf {\bibinfo {volume} {84}},\
  \bibinfo {pages} {5304} (\bibinfo {year} {2000})}\BibitemShut {NoStop}%
\bibitem [{\citenamefont {Brecht}\ \emph {et~al.}(2014)\citenamefont {Brecht},
  \citenamefont {Eckstein}, \citenamefont {Ricken}, \citenamefont {Quiring},
  \citenamefont {Suche}, \citenamefont {Sansoni},\ and\ \citenamefont
  {Silberhorn}}]{Brecht2014}%
  \BibitemOpen
  \bibfield  {author} {\bibinfo {author} {\bibfnamefont {B.}~\bibnamefont
  {Brecht}}, \bibinfo {author} {\bibfnamefont {A.}~\bibnamefont {Eckstein}},
  \bibinfo {author} {\bibfnamefont {R.}~\bibnamefont {Ricken}}, \bibinfo
  {author} {\bibfnamefont {V.}~\bibnamefont {Quiring}}, \bibinfo {author}
  {\bibfnamefont {H.}~\bibnamefont {Suche}}, \bibinfo {author} {\bibfnamefont
  {L.}~\bibnamefont {Sansoni}}, \ and\ \bibinfo {author} {\bibfnamefont
  {C.}~\bibnamefont {Silberhorn}},\ }\href {\doibase
  10.1103/PhysRevA.90.030302} {\bibfield  {journal} {\bibinfo  {journal}
  {Physical Review A}\ }\textbf {\bibinfo {volume} {90}},\ \bibinfo {pages}
  {030302} (\bibinfo {year} {2014})}\BibitemShut {NoStop}%
\bibitem [{\citenamefont {Rohde}\ and\ \citenamefont
  {Ralph}(2006)}]{Rohde2006}%
  \BibitemOpen
  \bibfield  {author} {\bibinfo {author} {\bibfnamefont {P.~P.}\ \bibnamefont
  {Rohde}}\ and\ \bibinfo {author} {\bibfnamefont {T.~C.}\ \bibnamefont
  {Ralph}},\ }\href {\doibase 10.1080/09500340600578369} {\bibfield  {journal}
  {\bibinfo  {journal} {Journal of Modern Optics}\ }\textbf {\bibinfo {volume}
  {53}},\ \bibinfo {pages} {1589} (\bibinfo {year} {2006})}\BibitemShut
  {NoStop}%
\bibitem [{\citenamefont {Barnett}\ \emph {et~al.}(1998)\citenamefont
  {Barnett}, \citenamefont {Phillips},\ and\ \citenamefont
  {Pegg}}]{Barnett1998}%
  \BibitemOpen
  \bibfield  {author} {\bibinfo {author} {\bibfnamefont {S.}~\bibnamefont
  {Barnett}}, \bibinfo {author} {\bibfnamefont {L.}~\bibnamefont {Phillips}}, \
  and\ \bibinfo {author} {\bibfnamefont {D.}~\bibnamefont {Pegg}},\ }\href
  {http://www.sciencedirect.com/science/article/pii/S0030401898005112}
  {\bibfield  {journal} {\bibinfo  {journal} {Optics communications}\ ,\
  \bibinfo {pages} {45}} (\bibinfo {year} {1998})}\BibitemShut {NoStop}%
\bibitem [{\citenamefont {Avenhaus}\ \emph {et~al.}(2008)\citenamefont
  {Avenhaus}, \citenamefont {Coldenstrodt-Ronge}, \citenamefont {Laiho},
  \citenamefont {Mauerer}, \citenamefont {Walmsley},\ and\ \citenamefont
  {Silberhorn}}]{Avenhaus2008}%
  \BibitemOpen
  \bibfield  {author} {\bibinfo {author} {\bibfnamefont {M.}~\bibnamefont
  {Avenhaus}}, \bibinfo {author} {\bibfnamefont {H.}~\bibnamefont
  {Coldenstrodt-Ronge}}, \bibinfo {author} {\bibfnamefont {K.}~\bibnamefont
  {Laiho}}, \bibinfo {author} {\bibfnamefont {W.}~\bibnamefont {Mauerer}},
  \bibinfo {author} {\bibfnamefont {I.}~\bibnamefont {Walmsley}}, \ and\
  \bibinfo {author} {\bibfnamefont {C.}~\bibnamefont {Silberhorn}},\ }\href
  {\doibase 10.1103/PhysRevLett.101.053601} {\bibfield  {journal} {\bibinfo
  {journal} {Physical Review Letters}\ }\textbf {\bibinfo {volume} {101}},\
  \bibinfo {pages} {053601} (\bibinfo {year} {2008})}\BibitemShut {NoStop}%
\bibitem [{\citenamefont {Wasilewski}\ \emph {et~al.}(2006)\citenamefont
  {Wasilewski}, \citenamefont {Lvovsky}, \citenamefont {Banaszek},\ and\
  \citenamefont {Radzewicz}}]{Wasilewski2006}%
  \BibitemOpen
  \bibfield  {author} {\bibinfo {author} {\bibfnamefont {W.}~\bibnamefont
  {Wasilewski}}, \bibinfo {author} {\bibfnamefont {A.}~\bibnamefont {Lvovsky}},
  \bibinfo {author} {\bibfnamefont {K.}~\bibnamefont {Banaszek}}, \ and\
  \bibinfo {author} {\bibfnamefont {C.}~\bibnamefont {Radzewicz}},\ }\href
  {\doibase 10.1103/PhysRevA.73.063819} {\bibfield  {journal} {\bibinfo
  {journal} {Physical Review A}\ }\textbf {\bibinfo {volume} {73}},\ \bibinfo
  {pages} {063819} (\bibinfo {year} {2006})}\BibitemShut {NoStop}%
\bibitem [{\citenamefont {Christ}\ \emph {et~al.}(2011)\citenamefont {Christ},
  \citenamefont {Laiho}, \citenamefont {Eckstein}, \citenamefont {Cassemiro},\
  and\ \citenamefont {Silberhorn}}]{Christ2011}%
  \BibitemOpen
  \bibfield  {author} {\bibinfo {author} {\bibfnamefont {A.}~\bibnamefont
  {Christ}}, \bibinfo {author} {\bibfnamefont {K.}~\bibnamefont {Laiho}},
  \bibinfo {author} {\bibfnamefont {A.}~\bibnamefont {Eckstein}}, \bibinfo
  {author} {\bibfnamefont {K.~N.}\ \bibnamefont {Cassemiro}}, \ and\ \bibinfo
  {author} {\bibfnamefont {C.}~\bibnamefont {Silberhorn}},\ }\href {\doibase
  10.1088/1367-2630/13/3/033027} {\bibfield  {journal} {\bibinfo  {journal}
  {New Journal of Physics}\ }\textbf {\bibinfo {volume} {13}},\ \bibinfo
  {pages} {033027} (\bibinfo {year} {2011})}\BibitemShut {NoStop}%
\bibitem [{\citenamefont {Patera}\ \emph {et~al.}(2009)\citenamefont {Patera},
  \citenamefont {Treps}, \citenamefont {Fabre},\ and\ \citenamefont
  {de~Valc\'{a}rcel}}]{Patera2009}%
  \BibitemOpen
  \bibfield  {author} {\bibinfo {author} {\bibfnamefont {G.}~\bibnamefont
  {Patera}}, \bibinfo {author} {\bibfnamefont {N.}~\bibnamefont {Treps}},
  \bibinfo {author} {\bibfnamefont {C.}~\bibnamefont {Fabre}}, \ and\ \bibinfo
  {author} {\bibfnamefont {G.~J.}\ \bibnamefont {de~Valc\'{a}rcel}},\ }\href
  {\doibase 10.1140/epjd/e2009-00299-9} {\bibfield  {journal} {\bibinfo
  {journal} {The European Physical Journal D}\ }\textbf {\bibinfo {volume}
  {56}},\ \bibinfo {pages} {123} (\bibinfo {year} {2009})}\BibitemShut
  {NoStop}%
\bibitem [{\citenamefont {Gradsshteyn}\ and\ \citenamefont
  {Ryzhik}(2007)}]{Gradsshteyn2007}%
  \BibitemOpen
  \bibfield  {author} {\bibinfo {author} {\bibfnamefont {I.}~\bibnamefont
  {Gradsshteyn}}\ and\ \bibinfo {author} {\bibfnamefont {I.}~\bibnamefont
  {Ryzhik}},\ }\href@noop {} {\emph {\bibinfo {title} {{Table of Integrals,
  Series, and Products}}}},\ \bibinfo {edition} {7th}\ ed.\ (\bibinfo
  {publisher} {Academic Press},\ \bibinfo {year} {2007})\BibitemShut {NoStop}%
\bibitem [{\citenamefont {Averchenko}\ \emph {et~al.}(2010)\citenamefont
  {Averchenko}, \citenamefont {Golubev}, \citenamefont {Fabre},\ and\
  \citenamefont {Treps}}]{Averchenko2010}%
  \BibitemOpen
  \bibfield  {author} {\bibinfo {author} {\bibfnamefont {V.}~\bibnamefont
  {Averchenko}}, \bibinfo {author} {\bibfnamefont {Y.}~\bibnamefont {Golubev}},
  \bibinfo {author} {\bibfnamefont {C.}~\bibnamefont {Fabre}}, \ and\ \bibinfo
  {author} {\bibfnamefont {N.}~\bibnamefont {Treps}},\ }\href {\doibase
  10.1140/epjd/e2010-00280-7} {\bibfield  {journal} {\bibinfo  {journal} {The
  European Physical Journal D}\ }\textbf {\bibinfo {volume} {61}},\ \bibinfo
  {pages} {207} (\bibinfo {year} {2010})}\BibitemShut {NoStop}%
\bibitem [{\citenamefont {Jiang}\ \emph {et~al.}(2012)\citenamefont {Jiang},
  \citenamefont {Treps},\ and\ \citenamefont {Fabre}}]{Jiang2012}%
  \BibitemOpen
  \bibfield  {author} {\bibinfo {author} {\bibfnamefont {S.}~\bibnamefont
  {Jiang}}, \bibinfo {author} {\bibfnamefont {N.}~\bibnamefont {Treps}}, \ and\
  \bibinfo {author} {\bibfnamefont {C.}~\bibnamefont {Fabre}},\ }\href
  {\doibase 10.1088/1367-2630/14/4/043006} {\bibfield  {journal} {\bibinfo
  {journal} {New Journal of Physics}\ }\textbf {\bibinfo {volume} {14}},\
  \bibinfo {pages} {043006} (\bibinfo {year} {2012})}\BibitemShut {NoStop}%
\bibitem [{\citenamefont {Grice}\ \emph {et~al.}(2001)\citenamefont {Grice},
  \citenamefont {U'Ren},\ and\ \citenamefont {Walmsley}}]{Grice2001}%
  \BibitemOpen
  \bibfield  {author} {\bibinfo {author} {\bibfnamefont {W.}~\bibnamefont
  {Grice}}, \bibinfo {author} {\bibfnamefont {A.}~\bibnamefont {U'Ren}}, \ and\
  \bibinfo {author} {\bibfnamefont {I.}~\bibnamefont {Walmsley}},\ }\href
  {\doibase 10.1103/PhysRevA.64.063815} {\bibfield  {journal} {\bibinfo
  {journal} {Physical Review A}\ }\textbf {\bibinfo {volume} {64}},\ \bibinfo
  {pages} {063815} (\bibinfo {year} {2001})}\BibitemShut {NoStop}%
\end{thebibliography}%

\appendix

\section{General photon subtraction kernel}\L{Kgeneral}

Splitting of a photon (\ref{out_AB}), its filtering (\ref{filtering}) and detection (\ref{P_t}) leads to the following conditioned state (non-normalized) of the signal light
	\begin{align}
	& \hat\rho^-_t \propto \iint \ud\w \ud\w' \; S_t(\w,\w') \; \hat a(\w') \hat\rho \hat a^\dag(\w)
	\end{align}
with the subtraction kernel
	\begin{align}
	\nn & S_t(\w,\w') \\
	&= \iint \ud\v \ud\v' R(\v',\w') R^*(\v,\w) F(\v') F^*(\v) \G(\v-\v') e^{i(\v-\v')t} 
	\L{K_full}
	\end{align}
Here $\G(\w) = \int \g(t) \, e^{i \w t} \ud\t/2\pi$ is a Fourier transform of the the detector temporal averaging function.

The conditioned state depends on the instant $t$ of the photon detection. The kernel is Hermitian and its eigen-decomposition gives spectral profiles of subtraction modes $\{v_j(\w)\}$ and corresponding subtraction efficiencies $\{\s_j\}$. Time-dependent phase factor $e^{i(\nu-\nu')t}$ shows that in time domain subtraction modes are centered at the instant of the detector click. Further we omit this term. Then eigen-decomposition formally reads:
	\begin{align}
	& S(\w,\w') = \sum\limits_j \s_j v_j(\w)  v_j^*(\w')
	\end{align}

\subsection{Beamsplitter with filtering}\L{app:Gauss-decomp}

For the beamsplitter $R(\w,\w') = r \;\dd(\w-\w')$ and one gets
	\begin{align}
	S(\w,\w') = r^2  F^*(\w) F(\w') \G(\w-\w')
	\end{align}
Eigen-decomposition of the kernel can be performed analytically in the Gaussian approximation of spectral filter with the bandwidth $\w_f$ and Gaussian averaging function of the photodetector $ \g(t) = \exp(-\t^2/\t_d^2)$ with the response time $\t_d$. In the frequency domain one has
	 \begin{align}
	& F(\w) = \exp(-\w^2/2\,\w_f^2),\L{F_func}\\ 
	& \G(\w) \propto \exp(-\t_d^2 \w^2/4)
	\end{align}
Then subtraction modes are given by the Hermite-Gaussian functions
	\begin{align}
	& v_j(\w) \propto \text{H}_j(\tau\w) \; \exp(-\t^2 \w^2/2)
	\L{HG}
	\end{align}
with the following characteristic temporal width
	\begin{align}
	& \t = \w_f^{-1}\sqrt[4]{1+\t_d^2\w_f^2}
	\end{align}
Subtraction efficiencies are given by
	\begin{align}
	& \s_j = \sigma \frac{\e^{2j-1}}{(1+\sqrt{1+\e^2})^{2j-1}}, \; (j\geq0)
	\end{align}
where $\s=r^2$ and $\e = \t_d \w_f$.
The number of subtraction modes is characterized with a Schmidt number $K$ given by:
	\begin{align}
	& K = \frac{(\sum \s_j)^2}{\sum \s_j^2} = \sqrt{1+\t_d^2 \w_f^2}
	\L{Sch_BS}
	\end{align}
Using above expressions one can distinguish the following regimes of the photon subtraction depending on relation between filter and photodetector parameters
	\begin{itemize}
	\item
	single-mode (fast detector and narrow filter $\t_d \w_f \ll 1$)
	\begin{align}
	& K=1 \quad \text{and} \quad v(\w) = F(\w)
	\end{align}
	The only subtraction mode is defined by the filtering function;
	\item
	multi-mode (slow detector and broad filter
	$\t_d \w_f \gg 1$)
		\begin{align}
		& K = \t_d\w_f \gg 1 \quad \text{and} \quad \t = \sqrt{\t_d/\w_f}
		\end{align}
	\end{itemize}
To have an additional intuition let us consider the photon subtraction in the temporal domain. Using beamsplitter transformation (\ref{out_AB}) one arrives at the wave-function of the signal and split beams after the beamsplitter and filtering (\ref{filtering}):
	\begin{align}
	\int \ud t \; \hat b^\dag(t) \(\int \ud t' \; f(t'-t) \,  \, \hat a(t')\) |\text{in}\> \otimes |0\>
	\L{BS_AB}
	\end{align}
where $f(t) = (2\pi)^{-1} \int\ud\w \; F(\w) e^{-i\w t}$ is a response of the filter in the temporal domain. Then instant detection of a photon at time $t$ in the split beam projects signal beam on the following state
	\begin{align}
	& |\text{out}\> \propto \hat a_t \inn,\\
	& \text{where:} \quad \hat a_t \propto \int \ud t' \; f(t'-t) \,  \, \hat a(t')
	\end{align}
The expression means that a photon is subtracted from a spectro-temporal mode defined by the filter. When the photon detection has a finite temporal resolution described by the expression (\ref{P_t}) the conditioned signal state is mixed
	\begin{align}
	\hat\rho \propto \int\ud\t \, \g(\t-t) \; \hat a_\t \; \inn \<\text{in}| \; \hat a_\t^\+ 
	\end{align}
One notes that subtraction modes, defined in this way,  are not orthogonal for different $\t$. Orthogonal modes are obtained via eigen-decomposition of the subtraction kernel performed above.

\subsection{Weak parametric up-conversion}

The characteristic time-scales of the parametric up-conversion are defined by the duration of gate pulses and inversed bandwidth of the up-conversion which are typically much shorter than the temporal resolution of the detector. Assuming also that the spectral filtering of the up-converted photons is moderate it is reasonable to treat photodetector as a slow one and use in the expression (\ref{K_full}) the approximation $\G(\nu-\nu') \approx \dd(\nu-\nu')$. One gets the following subtraction kernel
	\begin{align}
	& S(\w,\w') = \int \ud\nu R^*(\nu,\w) R(\nu,\w') |F(\nu)|^2
	\L{S_up}
	\end{align}

To get explicit results consider the up-conversion in a degenerate type-I configuration in a bulk crystal \cite{Averchenko2014}. Assuming collinear interaction of signal and gate fields the parametric kernel  is the product of the spectral distribution of gate field and phase-matching function, i.e. $R(\w,\w') \propto \a(\w-\w') \Phi(\w)$. Under the chosen conditions the phase-matching function depends only on the frequency of the up-converted photon. Then the subtraction kernel reads
	\begin{align}
	\nn & S(\w,\w') \\
	& \propto \int \ud\nu \a(\nu-\w') \a^*(\nu-\w) \Phi^2(\nu) |F(\nu)|^2
	\end{align}
One sees that reducing phase-matched bandwidth of the up-conversion (for example, choosing longer nonlinear medium) and performing narrowband filtering of the converted photon, so that  $\Phi^2(\nu) |F(\nu)|^2 \propto \dd(\nu)$, one can reach the single-mode regime where the only subtraction mode is defined by the gate pulse: $S(\w,\w') \propto \a^*(\w) \a(\w')$.

To get analytical results let us consider up-conversion with the Gaussian gate pulses:
	\begin{align}
	& \a(\w) \propto \exp(-\t_g^2\w^2/2)
	\end{align}
Further, using the relation $\sinc(x) \approx e^{-\gamma x^2}, \; \g\approx 0.193$ (see, for example, \cite{Grice2001}),  we approximate the phase-matching function with the Gaussian of the same width $\w_\text{ph}$	
	\begin{align}
	& \Phi(\w) \approx \exp(-\w^2/2\w_\text{ph}^2)
	\end{align}
For the type-I degenerate up-conversion the phase-matching width reads: $\w_\text{ph} \approx 1/\sqrt{\gamma/2} (v_c^{-1}-v_s^{-1})l$, where $l$ is the crystal length and $v_s, v_c$ are group velocities of the signal and converted photons.
We also assume Gaussian spectral filter in the form (\ref{F_func}). Then the subtraction modes are Hermite-Gaussian functions (\ref{HG}) with the duration	
	\begin{align}
	& \t = \t_g \sqrt[4]{1-\frac{1}{1+\t_g^{-2} (\w_\text{ph}^{-2}+ \w_f^{-2})}}
	\end{align}
Subtraction efficiencies are
	\begin{align}
	& \s_j = \s \frac{\e^{2j-1}}{(1+\sqrt{1-\e^2})^{2j-1}},
	\end{align}
where $\s=2\sqrt{\pi}|C|^2/\t_g$ and $\e=1/\sqrt{1+\t_g^{-2} (\w_\text{ph}^{-2}+ \w_f^{-2})}$.
The number of subtraction  modes reads
	\begin{align}
	& K= \sqrt{1+\frac{\t_g^2}{\w_\text{ph}^{-2}+\w_f^{-2}}}
	\end{align}
Firstly, the single mode subtraction is achieved for narrowband up-conversion with long crystal. Secondly, filtering of the up-converted photon can further reduce Schmidt number of the process. Then the only subtraction mode is defined by the gate pulse.

\section{Subtracting a single-photon from a squeezed frequency comb}
\label{sec:appendix_comb}

We detail the changes from the subtraction of a single photon from an individually squeezed pulse to a train of squeezed correlated pulses, so-called squeezed frequency comb.
An optical cavity enhances the parametric down-conversion and leads to correlations between successively generated signal pulses.
As a result a squeezed state is no longer concentrated in a single pulse, but distributed over a train of pulses.
It is shown in \cite{Jiang2012} that an individual mode of squeezing in the temporal domain is a continuous train of pulses.
A temporal mode is thus defined by the shape $\psi_k(t)$ of pulses that constitute the train of period $T_0$ and a phase-shift from one pulse to the next one $\theta$ ($\theta=-\pi \ldots \pi$):
	\begin{align}
	u_k(\theta,t) = \frac{1}{\sqrt{2 \pi}} \sum\limits_{l=-\infty}^{\infty} \psi_k(t-l T_0) e^{i l \theta}
	\end{align}
Applying the Fourier transformation one can show that this mode constitutes a comb in the spectral domain.
The comb envelope being the Fourier transform of $\psi_k(t)$.
The squeezed vacuum quantum state embedded in the comb can be described by the following transformation of the comb amplitude $\hat a_k(\tt)$:
	\begin{align}
	\hat a_k(\tt) = C_k(\tt) \; \hat a_k^\text{vac}(\tt) + S_k(\tt) \; \hat a_k^\text{vac}(-\tt)
	\end{align}
where the coefficients $C_k(\tt), S_k(\tt)$ are defined by the parametric down-conversion taking place in the cavity.
Rigorously speaking the transformation describes squeezing if $\tt=0$ and two-mode squeezing when $\tt \neq 0$.
However it will not affect further conclusions.

It is then possible to consider the properties of an individual pulse that constitutes the correlated train.
We pick a pulse labelled $l$ that have a profile corresponding to index $k$. It is showed in \cite{Jiang2012} that its amplitude reads:
	\begin{equation}
	\hat a_{k,l} = \int_{-\pi}^{\pi} \frac{\ud\tt}{\sqrt{2\pi}} \hat a_k(\tt) e^{i l \tt}
	\end{equation}
So that the field is a sum of the fields of squeezed modes.
It is also shown in \cite{Averchenko2010,Jiang2012} that there are inter-pulse correlations that decays exponentially with the distance between pulses.
To understand single-photon subtraction from such light, it is more convenient to analyze it in the temporal domain. An event of a fast single-photon detector (faster than period of pulses) heralds the subtraction of a photon from an individual pulse of light that we denote with the number $l=0$.
This physical condition formally means that the subtraction modes can be expanded as in expression \ref{eq:sub_expansion} :
	\begin{align}
	\hat s_j & \approx \sum_k c_{jk} \hat a_{k,l=0} \\
	& = \sum_k c_{jk} \int_{-\pi}^{\pi} \frac{\ud\tt}{\sqrt{2\pi}} \hat a_k(\tt)
	\end{align}
The first expression represents an expansion over the modes that constitute a single pulse while the second one gives an expansion over the squeezed modes of the multimode squeezed state.
In particular, one notices that the single-photon subtraction is not selective with respect to the $\theta$ parameter.

The subtraction probability and the purity of the conditioned quantum state are then given by expressions similar to \ref{P-1} and \ref{eq:multimode_purity}, respectively :
	\begin{align}
    	& P = \sum\limits_{j,k} \s_j |c_{jk}|^2 n_{k,0},\\
    	& \pi = \frac{\sum\limits_{j,j'} \s_j \s_{j'} \left|\sum\limits_k c_{jk} c_{j'k}^* n_{k,0}\right|^2}{P^2}
	\end{align}
Those expressions are similar to the ones obtained in section \ref{pulses}.
The only change lies in the photon number $n_{k,0}$ :
\begin{equation}
n_{k,0} = \<\hat a_{k,l=0}^\dag \hat a_{k,l=0}\> = \int \<\hat a_{k}^\dag(\tt) \hat a_{k}(\tt)\>\ud\tt/2\pi
\end{equation}
It is now the number of photons per single pulse, not per squeezed mode.

However if we consider the state of an individual mode of squeezing after single-photon subtraction it will be different from section \ref{pulses}.
Indeed, we have showed that the conditioned state within a squeezed mode is defined by the probability that the detected photon belongs to the mode.
In other words how mode-selective the single-photon subtraction is. 
If the photon is subtracted from an individual pulse, the subtraction is not selective with respect to the phase shift parameter $\tt$.
Therefore, the conditional probability that the detected photon belongs to a specific squeezed mode within a range $\D\theta$  is smaller by an amount $\pi/\D\theta \gg 1$.
Consequently, the Wigner function of the state of a squeezed mode will not possesses any negativity after the single-photon subtraction.

This remark also concerns the mode of an heralded single-photon state.
In the week squeezing approximation one can find a mode with the highest subtraction probability that will host a single photon state and so display negative values in its Wigner function.

\end{document}